\documentclass[useAMS,usenatbib]{mn2e}
\usepackage{txfonts}
\usepackage{graphicx}
\usepackage{natbib}
\usepackage{amssymb}

\def\del#1{{}}

\defcitealias{2007MNRAS...378..385P}{Paper I}
\defcitealias{2007PfrommerII}{Paper II}

\del{
\bibitem[\protect\citeauthoryear{{Miniati}, {Ryu}, {Kang} \& {Jones}}{{Miniati}
  et~al.}{2001a}]{2001ApJ...559...59M}
{Miniati} F.,  {Ryu} D.,  {Kang} H.,    {Jones} T.~W.,  2001a, \apj, 559, 59
\bibitem[\protect\citeauthoryear{{Miniati}, {Jones}, {Kang} \& {Ryu}}{{Miniati}
  et~al.}{2001b}]{2001ApJ...562..233M}
{Miniati} F.,  {Jones} T.~W.,  {Kang} H.,    {Ryu} D.,  2001b, \apj, 562, 233
\bibitem[\protect\citeauthoryear{{Pfrommer}, {En{\ss}lin} \&
  {Springel}}{{Pfrommer} et~al.}{2007b}]{2007PfrommerII}
{Pfrommer} C.,  {En{\ss}lin} T.~A.,    {Springel} 2007b, ArXiv:0707.1707
\bibitem[\protect\citeauthoryear{{Pfrommer}, {En{\ss}lin}, {Springel},
  {Jubelgas} \& {Dolag}}{{Pfrommer} et~al.}{2007a}]{2007MNRAS...378..385P}
{Pfrommer} C.,  {En{\ss}lin} T.~A.,  {Springel} V.,  {Jubelgas} M.,    {Dolag}
  K.,  2007a, \mnras, 378, 385
}

\newcommand{\apj}{ApJ}

\newcommand{\mnras}{MNRAS}

\newcommand{\dd}{\mathrm{d}}
\newcommand{\bra}{\langle}
\newcommand{\ket}{\rangle}
\newcommand{\ltsima}{$\; \buildrel < \over \sim \;$}
\newcommand{\lsim}{\lower.5ex\hbox{\ltsima}}
\newcommand{\gtsima}{$\; \buildrel > \over \sim \;$}
\newcommand{\gsim}{\lower.5ex\hbox{\gtsima}}

\newcommand{\e}{{\rm e}}
\newcommand{\p}{{\rm p}}
\newcommand{\inj}{{\rm inj}}

\newcommand{\CR}{{\rm CR}}
\newcommand{\CRe}{\rmn{CRe}}
\renewcommand{\th}{{\rm th}}

\newcommand{\IC}{\rmn{IC}}

\newcommand{\M}{{\mathcal M}}
\newcommand{\F}{{\mathcal F}}
\renewcommand{\L}{{\mathcal L}}

\newcommand{\vir}{\mathrm{vir}}
\newcommand{\eps}{\varepsilon}

\newcommand{\vel}{\upsilon}

\title[Cosmic rays in clusters of galaxies -- III. Comparison to observations]
{Simulating cosmic rays in clusters of galaxies -- III. Non-thermal scaling
  relations and comparison to observations}

\author[C.~Pfrommer]
  {Christoph~Pfrommer\thanks{e-mail: pfrommer@cita.utoronto.ca} \\ 
  Canadian Institute for Theoretical Astrophysics, University of Toronto,
  60 St. George Street, Toronto, Ontario, M5S 3H8, Canada }

\begin{document}
\pagerange{\pageref{firstpage}--\pageref{lastpage}} \pubyear{2003}
\maketitle
\label{firstpage}

\begin{abstract}
  Complementary views of galaxy clusters in the radio synchrotron, hard X-ray
  inverse Compton, and high-energy $\gamma$-ray regimes are critical in
  calibrating them as high-precision cosmological probes.  We present
  predictions for scaling relations between cluster mass and these non-thermal
  observables.  To this end, we use high-resolution simulations of a sample of
  galaxy clusters spanning a mass range of almost two orders of magnitudes, and
  follow self-consistent cosmic ray physics on top of the radiative
  hydrodynamics.  We model relativistic electrons that are accelerated at
  cosmological structure formation shocks and those that are produced in
  hadronic interactions of cosmic rays with ambient gas protons.  Calibrating
  the magnetic fields of our model with Faraday rotation measurements, the
  synchrotron emission of our relativistic electron populations matches the
  radio synchrotron luminosities and morphologies of observed giant radio halos
  and mini-halos surprisingly well. Using the complete sample of the brightest
  X-ray clusters observed by ROSAT in combination with our $\gamma$-ray scaling
  relation, we predict GLAST that will detect about ten clusters allowing for
  Eddington bias due to the scatter in the scaling relation.  The expected
  brightest $\gamma$-ray clusters are Ophiuchus, Fornax, Coma, A3627, Perseus,
  and Centaurus. The high-energy $\gamma$-ray emission above 100~MeV is
  dominated by pion decays resulting from hadronic cosmic ray interactions. We
  provide an absolute lower flux limit for the $\gamma$-ray emission of Coma in
  the hadronic model which can be made tighter for magnetic field values
  derived from rotation measurements to match the GLAST sensitivity, providing
  thus a unique test for the possible hadronic origin of radio halos.  Our
  predicted hard X-ray emission, due to inverse Compton emission of shock
  accelerated and hadronically produced relativistic electrons, falls short of
  the detections in Coma and Perseus by a factor of 50. This casts doubts on
  inverse Compton interpretation and reinforces the known discrepancy of
  magnetic field estimates from Faraday rotation measurements and those
  obtained by combining synchrotron and inverse Compton emission.
\end{abstract}

\begin{keywords}
  galaxies: cluster: general, cosmic rays, magnetic fields, radiation
  mechanisms: non-thermal
\end{keywords}

\section{Introduction}

Previously, it has been assumed that galaxy clusters are sufficiently well
described by their mass which was thought to be largely independent of the
complex astrophysical processes taking place in the intra-cluster medium (ICM)
such as star formation and different kinds of feedback processes.
High-resolution {\em XMM-Newton} and {\em Chandra} X-ray observations taught us
in the last years that this over-simplified paradigm needs to be modified.
Even ostensibly `relaxed' clusters reveal a richness of substructure with
substantial small-scale variation in temperature, metallicity, and surface
brightness.

This raises the question if high-precision cosmology will principally be
possible using clusters.  Clearly, we need to understand how non-equilibrium
processes that lead to cosmic ray populations and turbulence impact on the
thermal X-ray emission and Sunyaev-Zel'dovich effect. This forces us to
explore complementary observational windows to clusters such as non-thermal
emission that can potentially elucidate the otherwise invisible
non-equilibrium processes. The upcoming generation of low-frequency radio,
hard X-ray, and $\gamma$-ray instruments open up the extragalactic sky in
unexplored wavelength ranges \citep[cf.][ for a compilation of these
experiments]{2007PfrommerII}.  Suitably combining radio synchrotron radiation,
inverse Compton emission in the hard X-ray regime, and high-energy
$\gamma$-ray emission will enable us to estimate the cosmic ray pressure
contribution and provide us with clues to the dynamical state of a cluster.
This will allow us to construct a `gold sample' for cosmology using
information on the dynamical cluster activity that is orthogonal to the
thermal cluster observables. Additionally, these non-thermal observations have
the potential to improve our knowledge of diffusive shock acceleration, large
scale magnetic fields, and turbulence.

Of the possible non-thermal emission bands, only the diffuse large-scale radio
synchrotron emission of clusters has been unambiguously detected so far.
Generally these radio phenomena can be divided into two categories that differ
morphologically, in their degree of polarisation, as well as in their
characteristic emission regions with respect to the cluster halo.  The
large-scale `radio relic' or `radio gischt' emission
\citep{2004rcfg.proc..335K}, that has a high degree of polarisation, is
irregularly shaped and occurs at peripheral cluster regions, can be attributed
to merging or accretion shock waves as proposed by \citet{1998A&A...332..395E}.
Prominent examples for large scale `radio relic' emission have been observed
in Abell~3667 \citep{1997MNRAS.290..577R}, Abell~3376
\citep{2006Sci...314..791B}, and Abell~2256 \citep{2006AJ....131.2900C}.  In
contrast, `cluster radio halos' show a coherently large diffuse radio
emission that is centred on the cluster, resemble the underlying thermal
bremsstrahlung emission in X-rays, are unpolarised, and show spectral index
variations that are amplified in the peripheral regions of the extended radio
emitting regions.  These radio halo phenomena can be furthermore subdivided
into Mpc-sized `giant radio halos' that are associated with merging clusters
and `radio mini-halos' that are observed in a few cool core clusters and
have a smaller extent of a few hundreds of kpc.  Prominent examples for
`giant radio halos' can be obtained from \citet{1999NewA....4..141G} and
include the Coma cluster \citep{1989Natur.341..720K, 1997A&A...321...55D} and
the galaxy cluster 1E 0657-56 \citep{2000ApJ...544..686L}. Prominent `radio
mini halos' are observed in the Perseus cluster \citep{1990MNRAS.246..477P} or
in RX~J1347.5-1145 \citep{2007A&A...470L..25G}.

Previously, there have been two models suggested that are able to explain
`cluster radio halos'. (1) {\em Re-acceleration processes} of `mildly'
relativistic electrons ($\gamma\simeq 100-300$) that are being injected over
cosmological timescales into the ICM by sources like radio galaxies, merger
shocks, or galactic winds can provide an efficient supply of highly-energetic
relativistic electrons.  Owing to their long lifetimes of a few times $10^9$
years these `mildly' relativistic electrons can accumulate within the ICM
\citep{2002mpgc.book....1S}, until they experience continuous in-situ
acceleration either via interactions with magneto-hydrodynamic waves, or
through turbulent spectra \citep{1977ApJ...212....1J, 1987A&A...182...21S,
  2001MNRAS.320..365B, 2002ApJ...577..658O, 2004MNRAS.350.1174B,
  2004A&A...417....1G, 2007MNRAS.378..245B}.  (2) {\em Hadronic interactions
  of relativistic protons} with ambient gas protons produce pions which decay
successively into secondary electrons, neutrinos and $\gamma$-rays. These
secondary relativistic electrons and positrons can emit a halo of radio
synchrotron emission in the presence of ubiquitous intra-cluster magnetic
fields \citep{1980ApJ...239L..93D, 1982AJ.....87.1266V, 1999APh....12..169B,
  2000A&A...362..151D, 2001ApJ...562..233M, 2003A&A...407L..73P,
  2004A&A...413...17P, 2004MNRAS.352...76P} as well as inverse Compton
emission by scattering photons from the cosmic microwave background into the
hard X-ray and $\gamma$-regime.  In our companion paper \citep[][ hereafter
Paper II]{2007PfrommerII}, we suggest a modification of the latter model that
is motivated by our high-resolution cluster simulations and cures the
weaknesses of the original model. We find, that our simulated giant radio
halos are dominated in the centre by secondary synchrotron emission with a
transition to the radio synchrotron radiation emitted from shock-accelerated
electrons in the cluster periphery.  This explains the extended radio emission
found in merging clusters, while it is more centrally concentrated in relaxed
cool core clusters. Varying spectral index distributions preferably in the
cluster periphery \citep{2004JKAS...37..315F} support this picture. The
characterisation of quantities related to cosmic rays in clusters can be found
in our first companion paper that studies the interplay of thermal gas and
cosmic rays and their effect on thermal cluster observables such as X-ray
emission and the Sunyaev-Zel'dovich effect \citep[][ hereafter Paper
I]{2007MNRAS...378..385P}.

The outline of the paper is as follows. Section~\ref{sec:methodology} describes
our general methodology, presents our cluster sample, and the different
simulated physical processes.  In Sect.~\ref{sec:results}, we present the
results on the cluster scaling relations for non-thermal observables as well as
$\gamma$-ray flux and luminosity functions.  These are compared to observations
and finally critically discussed in Sect.~\ref{sec:discussion}.

\section{Methodology}
\label{sec:methodology}

\subsection{General procedure}

We have performed high-resolution hydrodynamic simulations of the formation of
14 galaxy clusters. The clusters span a mass range from $5 \times 10^{13}\,
h^{-1}\, \rmn{M}_\odot$ to $2 \times 10^{15}\, h^{-1}\, \rmn{M}_\odot$ and show
a variety of dynamical states ranging from relaxed cool core clusters to
violent merging clusters (cf. Table~\ref{tab:sample}).  Our simulations
dynamically evolve dissipative gas physics including radiative cooling, star
formation, and supernova feedback. We identify the strength of structure
formation shock waves on-the-fly in our simulations and measure the shock Mach
number that is defined by the ratio of shock velocity to pre-shock sound
velocity, $\M=\vel_\rmn{shock} / c_\rmn{sound}$ \citep{2006MNRAS.367..113P}.
On top of this, we self-consistently follow cosmic ray (CR) physics including
adiabatic CR transport processes, injection by supernovae and cosmological
structure formation shocks, as well as CR thermalization by Coulomb interaction
and catastrophic losses by hadronic interactions \citep{2007A&A...473...41E,
  2006...Jubelgas}. In our post-processing, we model relativistic electrons
that are accelerated at cosmological structure formation shocks and those that
are produced in hadronic interactions of cosmic rays with ambient gas protons.
This approach is justified since these electrons do not modify the
hydrodynamics of the gas owing to their negligible pressure contribution. We
compute the stationary relativistic electron spectrum that is obtained by
balancing the mentioned injection mechanisms with the synchrotron and inverse
Compton cooling processes.  Details of our modelling can be found in
\citetalias{2007PfrommerII}.  Both populations of relativistic electrons emit a
morphologically distinguishable radio synchrotron radiation as well as inverse
Compton emission due to up-scattering of photons of the cosmic microwave
background (CMB) into the hard X-ray and $\gamma$-ray regime. At energies
larger than $100$~MeV, we expect additionally $\gamma$-ray emission from
decaying pions that are produced in hadronic CR interactions.  While the
emission of the shock accelerated {\em primary electrons} is amorphous and
peripheral as observed in radio relics, the hadronically produced {\em
  secondary electrons} show a centrally concentrated emission characteristic
that resembles that of the central parts of observed radio halos.

In this paper, we concentrate on three observationally motivated wave-bands.
(1) Radio synchrotron emission at $1.4$~GHz, (2) non-thermal hard X-ray
emission at energies $E_\gamma>10$~keV, and (3) $\gamma$-ray emission at
energies $E_\gamma>100$~MeV. We study the contribution of the different
emission components to the total cluster luminosity in each of these bands,
derive cluster scaling relations, and study their dependence on the simulated
physics and adopted parametrisation of the magnetic field.  The radio
synchrotron scaling relation is then compared to the observed sample of giant
radio halos and radio mini-halos. Using cluster masses from the complete sample
of the X-ray brightest clusters (HIFLUGCS, the HIghest X-ray FLUx Galaxy
Cluster Sample, \citet{2002ApJ...567..716R}), we construct luminosity and flux
functions for the hard X-ray and $\gamma$-ray band. This allows us to identify
the brightest clusters in the hard X-ray and $\gamma$-ray sky and predict the
cluster sample to be seen by GLAST.

\subsection{Adopted cosmology and cluster sample}

We provide only a short overview over the simulations and our cluster sample
for completeness while the simulation details can be found in
\citetalias{2007PfrommerII}.  Simulations were performed using the `concordance'
cosmological cold dark matter model with a cosmological constant
($\Lambda$CDM).  The cosmological parameters of our model are: $\Omega_\rmn{m}
= \Omega_\rmn{DM} + \Omega_\rmn{b} = 0.3$, $\Omega_\rmn{b} = 0.039$,
$\Omega_\Lambda = 0.7$, $h = 0.7$, $n = 1$, and $\sigma_8 = 0.9$.  Here,
$\Omega_\rmn{m}$ denotes the total matter density in units of the critical
density for geometrical closure today, $\rho_\rmn{crit} = 3 H_0^2 / (8 \upi
G)$. $\Omega_\rmn{b}$ and $\Omega_\Lambda$ denote the densities of baryons and
the cosmological constant at the present day. The Hubble constant at the
present day is parametrised as $H_0 = 100\,h \mbox{ km s}^{-1}
\mbox{Mpc}^{-1}$, while $n$ denotes the spectral index of the primordial
power-spectrum, and $\sigma_8$ is the {\em rms} linear mass fluctuation within
a sphere of radius $8\,h^{-1}$Mpc extrapolated to $z=0$.

\begin{table}
\caption{\scshape: Cluster sample}
\begin{tabular}{l l l l r r}
\hline
\hline
Cluster & sim.'s & dyn. state$^{(1)}$ & $M_{200}^{(2)}$ & $R_{200}^{(2)}$ & $kT_{200}^{(3)}$ \\
& & & [$h^{-1}\,\rmn{M}_\odot$] & [$h^{-1}\,$Mpc] & [keV] \\
\hline
1  & g8a  & CC    & $1.8\times 10^{15}$ & 2.0  & 13.1 \\
2  & g1a  & CC    & $1.3\times 10^{15}$ & 1.8  & 10.6 \\
3  & g72a & PostM & $1.1\times 10^{15}$ & 1.7  & 9.4  \\
4  & g51  & CC    & $1.1\times 10^{15}$ & 1.7  & 9.4  \\
                                                    
5  & g1b  & M     & $3.7\times 10^{14}$ & 1.2  & 4.7  \\
6  & g72b & M     & $1.5\times 10^{14}$ & 0.87 & 2.4  \\
7  & g1c  & M     & $1.4\times 10^{14}$ & 0.84 & 2.3  \\
8  & g8b  & M     & $1.0\times 10^{14}$ & 0.76 & 1.9  \\
9  & g1d  & M     & $9.2\times 10^{13}$ & 0.73 & 1.7  \\
                                                    
10 & g676 & CC    & $8.8\times 10^{13}$ & 0.72 & 1.7  \\
11 & g914 & CC    & $8.5\times 10^{13}$ & 0.71 & 1.6  \\
12 & g1e  & M     & $6.4\times 10^{13}$ & 0.65 & 1.3  \\
13 & g8c  & M     & $5.9\times 10^{13}$ & 0.63 & 1.3  \\
14 & g8d  & PreM  & $5.4\times 10^{13}$ & 0.61 & 1.2  \\
\hline
\end{tabular}   
\begin{quote} 
  {\scshape Notes:}\\
  (1) The dynamical state has been classified through a combined criterion
  invoking a merger tree study and the visual inspection of the X-ray
  brightness maps. The labels for the clusters are M--merger, PostM--post
  merger (slightly elongated X-ray contours, weak cool core region
  developing), PreM--pre-merger (sub-cluster already within the virial
  radius), CC--cool core cluster with extended cooling region (smooth X-ray
  profile).\\
  (2) The virial mass and radius are related by $M_\Delta(z) = \frac{4}{3}
  \pi\, \Delta\, \rho_\rmn{crit}(z) R_\Delta^3 $, where $\Delta=200$ denotes a
  multiple of the critical overdensity $\rho_\rmn{crit}(z) = 3 H (z)^2/ (8\pi
  G)$. \\  
  (3) The virial temperature is defined by $kT_\Delta = G M_\Delta \, \mu\,
  m_\p / (2 R_\Delta)$, where $\mu$ denotes the mean molecular weight.
\end{quote}
\label{tab:sample}
\end{table} 

We analysed the clusters with a halo-finder based on spherical overdensity
followed by a merger tree analysis in order to get the mass accretion history
of the main progenitor. We also produced projections of the X-ray emissivity at
redshift $z=0$ in order to get a visual impression of the cluster
morphology. The dynamical state of a cluster is defined by a combined
criterion: (i) if the cluster did not experience a major merger with a
progenitor mass ratio 1:3 or larger after $z=0.8$ (corresponding to a look-back
time of $\sim 5\, h^{-1}\,$Gyr) and (ii) if the visual impression of the
cluster's X-ray morphology is relaxed, it was defined to be a cool core
cluster.  The spherical overdensity definition of the virial mass of the
cluster is given by the material lying within a sphere centred on a local
density maximum, whose radial extend $R_\Delta$ is defined by the enclosed
threshold density condition $M (< R_\Delta) / (4 \pi R_\Delta^3 / 3) =
\rho_\rmn{thres}$. We chose the threshold density $\rho_\rmn{thres}(z) =
\Delta\, \rho_\rmn{crit} (z)$ to be a multiple $\Delta=200$ of the critical
density of the universe $\rho_\rmn{crit} (z) = 3 H (z)^2/ (8\pi G)$. We assume a
constant $\Delta=200$ although some treatments employ a time-varying $\Delta$
in cosmologies with $\Omega_\rmn{m} \ne 1$ \citep{1996MNRAS.282..263E}. In the
reminder of the paper, we use the terminology $R_\rmn{vir}$ instead of
$R_{200}$.

\subsection{The models}
\label{sec:models}

\begin{table}
\caption{\scshape: Different physical processes in our simulations:}
\begin{center}
\begin{tabular}{l | c c c c c}
\hline
\hline
Simulated physics$^{(1)}$ & \multicolumn{5}{c}{simulation models$^{(1)}$:}\\
& S1 & & S2 & & S3 \\ 
\hline
thermal shock heating & \checkmark & & \checkmark & & \checkmark \\
radiative cooling     &            & & \checkmark & & \checkmark \\
star formation        &            & & \checkmark & & \checkmark \\
Coulomb CR losses     & \checkmark & & \checkmark & & \checkmark \\
hadronic CR losses    & \checkmark & & \checkmark & & \checkmark \\
shock-CRs             & \checkmark & & \checkmark & & \checkmark \\
supernova-CRs         &            & &            & & \checkmark \\
\hline
\end{tabular}   
\end{center}
\begin{quote} 
  {\scshape Notes:}\\
  (1) This table serves as an overview over our simulated models. The first
  column shows the simulated physics and the following three columns show our
  different simulation models with varying gas and cosmic ray physics. Model
  S1 models the thermal gas non-radiatively and includes CR physics, while the
  models S2 and S3 use radiative gas physics with different variants of
  CR physics.\\
\end{quote}
\label{tab:models}
\end{table} 

For each galaxy cluster we ran three different simulations with varying gas and
cosmic ray physics (cf.~Table~\ref{tab:models}).  The first set of simulations
used non-radiative gas physics only, i.e.~the gas is transported adiabatically
unless it experiences structure formation shock waves that supply the gas with
entropy and thermal pressure support.  Additionally we follow cosmic ray (CR)
physics including adiabatic CR transport processes, injection by cosmological
structure formation shocks with a Mach number dependent acceleration scheme, as
well as CR thermalization by Coulomb interaction and catastrophic losses by
hadronic interactions (model S1).  The second set of simulations follows the
radiative cooling of the gas, star formation, supernova feedback, and a
photo-ionising background \citepalias[details can be found
  in][]{2007MNRAS...378..385P}.  As before in model S1, we account for CR
acceleration at structure formation shocks and allow for all CR loss processes
(model S2). The last set of simulations additionally assumes that a constant
fraction $\zeta_\rmn{SN} = \eps_\rmn{CR,inj}/\eps_\rmn{diss} = 0.3$ of the
kinetic energy of a supernova ends up in the CR population (model S3), which is
motivated by TeV $\gamma$-ray observations of a supernova remnant that find an
energy fraction of $\zeta_\rmn{SN} \simeq 0.1 - 0.3$ when extrapolating the CR
distribution function \citep{2006Natur.439..695A}. We choose a maximum value
for this supernova energy efficiency owing to the large uncertainties and our
aim to bracket the realistic case with the two radiative CR simulations.

Since we have not run self-consistent magneto-{hy\-dro\-dy\-nam\-ical} (MHD)
simulations on top of the radiative gas and CR physics, we chose the following
model for the magnetic energy density to compute the synchrotron and inverse
Compton (IC) emission:
\begin{equation}
  \label{eq:magnetic_scaling}
  \eps_B = \eps_{B,0} \left(\frac{\eps_\rmn{th}}{\eps_\rmn{th,0}}\right)^{2
  \alpha_B},
\end{equation}
where the central magnetic energy density $\eps_{B,0}$ and $\alpha_B$ are free
parameters in our model, and $\eps_\rmn{th,0}$ denotes the thermal energy
density at the cluster centre. Rather than applying a scaling with the gas
density as non-radiative MHD simulations by \citet{1999A&A...348..351D,
  2001A&A...378..777D} suggest, we chose the energy density of the thermal gas.
This quantity is well behaved in the centres of clusters where current
cosmological radiative simulations, that do not include radio-mode feedback
from AGN, have an over-cooling problem which results in an overproduction of
the amount of stars, enhanced central gas densities, too small central
temperatures, and too strong central entropy plateaus compared to X-ray
observations.  Theoretically, the growth of magnetic field strength is
determined through turbulent dynamo processes that will saturate on a level
which is determined by the strength of the magnetic back-reaction
\citep[e.g.,][]{2003PhRvL..90x5003S, 2006PhPl...13e6501S} and is typically a
fraction of the turbulent energy density that itself should be related to the
thermal energy density, thus motivating our model theoretically.


\section{Results}
\label{sec:results}

\subsection{Radio synchrotron emission}
\label{sec:synchro}

\subsubsection{Simulated synchrotron scaling relations}
\label{sec:radio_scaling_relations}

\begin{figure}
\begin{center}
    \centering{\it \large Synchrotron emission ($\nu=1.4$~GHz):}
\resizebox{\hsize}{!}{\includegraphics{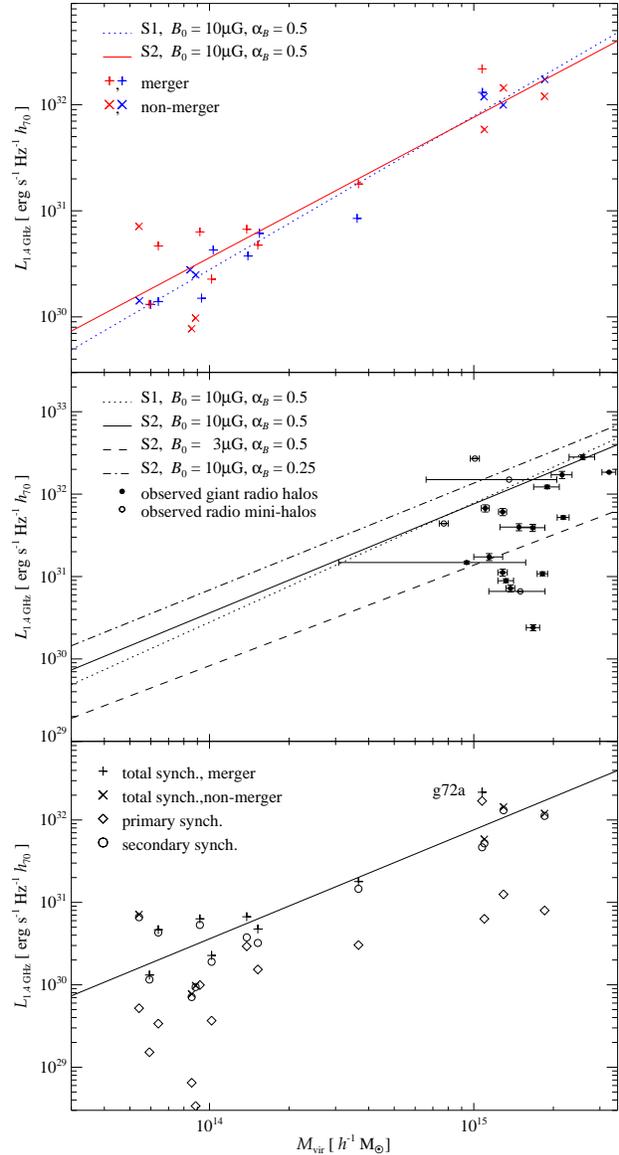}}
\end{center}
  \caption{Cluster scaling relations for the radio synchrotron luminosities at
    $\nu=1.4$~GHz.  The top panel shows the scatter of the individual clusters
    for our non-radiative (model S1) and radiative simulations (model S2). The
    middle panel shows the dependence of the scaling relations on the
    uncertainty in the magnetic field and simulated physics. The bottom panel
    shows the contribution of the individual emission components (primary,
    secondary radio synchrotron emission) to the total radio luminosities in
    our model S2 while assuming a central magnetic field strength of $B_0 =
    10\,\umu$G and an energy density scaling of $\alpha_B=0.5$.  }
  \label{fig:scalings_syn}
\end{figure}

In order to determine the cluster scaling relations for non-thermal
observables, we integrate the total surface brightness (composed of primary
and secondary emission components) within the virial radius of each cluster.
In our radiative simulations, we cut the region with $r < 0.025\, R_\rmn{vir}$
around the brightest central point-source that is caused by over-cooling gas of
the cD galaxy.  Since the modelled non-thermal emission processes reflect
active non-equilibrium structure formation processes, we expect a large
scatter in these scaling relations. Ideally, we would like to have a large
sample of independent clusters to obtain reliable measurements of the scaling
parameters. Thus, our limited sample will have larger uncertainties in the
derived parameters. Figure~\ref{fig:scalings_syn} shows our simulated
synchrotron scaling relations at $\nu=1.4$~GHz using the total radio
synchrotron luminosity within $R_\rmn{vir}$ of all clusters.  We note that the
radio emission volume is significantly enlarged for our merging clusters
mostly due to the larger contribution of primary radio emission in the cluster
outskirts \citepalias{2007PfrommerII}.  To simplify comparison with observed
giant radio halo samples, we additionally fit radio synchrotron scaling
relations for our subsample of eight merging clusters. The fit parameters for
our models with varying simulated physics and magnetic parameters can be found
in Table~\ref{tab:scaling}.  The following conclusions can be drawn.

\begin{table*}
\caption{\scshape Cluster scaling relations for non-thermal observables$^{(1)}$.}
\begin{tabular}{l c c  c c  c c  c c  c c }
\hline
\hline
&&
&\multicolumn{4}{c}{radio synchrotron:} & 
 \multicolumn{2}{c}{IC ($E_\rmn{IC} > 10$~keV):} & 
 \multicolumn{2}{c}{$\gamma$-rays ($E_\gamma > 100$~MeV):}\\
&&
&\multicolumn{2}{c}{all clusters} & \multicolumn{2}{c}{merging clusters}&&\\
\hline
model$^{(2)}$& $B_0^{(3)}$ & $\alpha_B^{(3)}$
& $L_{\nu,0}^{(4)}$ & $\beta_\nu$ & $L_{\nu,0}^{(4)}$ & $\beta_\nu$ 
& $\L_{\rmn{IC},0}^{(5)}$  & $\beta_\rmn{IC}$ 
& $\L_{\gamma,0}^{(6)}$ & $\beta_\gamma$ 
 \\
\hline
S1$^{(2)}$ & 10  & 0.5~~ & $0.78\pm0.05$ & $1.45\pm0.07$ & $0.80\pm0.12$ & $1.50\pm0.17$ & $2.95\pm0.26$ & $1.52\pm0.10$ & $7.85\pm0.58$ & $1.52\pm0.09$ \\ 
S2$^{(2)}$ & 10  & 0.5~~ & $0.76\pm0.12$ & $1.32\pm0.18$ & $1.27\pm0.26$ & $1.50\pm0.23$ & $1.66\pm0.22$ & $1.34\pm0.16$ & $5.46\pm0.74$ & $1.32\pm0.16$ \\
S2$^{(2)}$ & 10  & 0.25  & $1.36\pm0.19$ & $1.30\pm0.16$ & $2.59\pm0.41$ & $1.51\pm0.18$ &  &  &  &  \\      
S1$^{(2)}$ & ~~3 & 0.5~~ & $0.14\pm0.01$ & $1.40\pm0.09$ & $0.26\pm0.03$ & $1.46\pm0.20$ &  &  &  &  \\
S2$^{(2)}$ & ~~3 & 0.5~~ & $0.14\pm0.03$ & $1.22\pm0.23$ & $0.14\pm0.07$ & $1.46\pm0.30$ & $2.27\pm0.29$ & $1.33\pm0.15$ & $5.65\pm0.76$ & $1.32\pm0.16$ \\
S2$^{(2)}$ & ~~3 & 0.25  & $0.31\pm0.05$ & $1.27\pm0.18$ & $0.59\pm0.12$ & $1.49\pm0.21$ &  &  &  &  \\
S3$^{(2)}$ & 10  & 0.5~~ & $0.89\pm0.16$ & $1.07\pm0.21$ & $1.80\pm0.37$ & $1.31\pm0.23$ & $2.24\pm0.23$ & $1.24\pm0.12$ & $8.66\pm1.12$ & $1.17\pm0.15$ \\
\hline
\end{tabular}
\begin{quote} 
  {\scshape Notes:}\\ 
  (1) The cluster scaling relations for non-thermal observables are defined by
  $A = A_0\,M_{15}^{\beta}$, where $M_{15} = M_\vir / (10^{15} \rmn{M}_\odot /
  h)$ and the respective non-thermal luminosity is obtained by integrating over
  the virial region of the cluster within $R_{200}$ and applying a central
  cut around the brightest central point-source for radii $r < 0.025\,
  R_\rmn{vir}$. \\
  (2) The definition for our different models can be found in
  Table~\ref{tab:models}.\\
  (3) The definition for the parametrisation of the magnetic energy density is
  given by $\eps_B = \eps_{B,0}\,(\eps_\rmn{th}/\eps_\rmn{th,0})^{2 \alpha_B}$
  according to (\ref{eq:magnetic_scaling}) and $B_0$ is given in units of
  $\umu$G.\\  
  (4) The normalisation of the radio synchrotron scaling relations is given in
  units of $10^{32}\mbox{ erg s}^{-1}\mbox{ Hz}^{-1}\, h_{70}$. \\
  (5) The normalisation of the IC scaling relations ($E_\rmn{IC}>10$~keV) is
  given in units of $10^{49}\,\gamma\mbox{ s}^{-1}\, h_{70}$. \\
  (6) The normalisation of the $\gamma$-ray scaling relations
  ($E_\gamma>100$~MeV) is given in units of $10^{45}\,\gamma\mbox{ s}^{-1}\,
  h_{70}$. \\ 
\end{quote}
\label{tab:scaling}
\end{table*} 

{\bf Contributions of different emission components.} (1) The secondary
emission component is dominant for relaxing cool core (CC) clusters, and those
that only experience a minor merger. The primary component exceeds the
secondary one for major merging clusters by a factor of four as can be
seen in our large post-merging cluster g72a with $M\simeq 10^{15}
h^{-1}\,\rmn{M}_\odot$. (2) The secondary radio emission is remarkably similar
for our massive clusters while the scatter of the secondary emission increases
notably for our small clusters with $M\lesssim 2\times 10^{14}
h^{-1}\,\rmn{M}_\odot$. This is due to the property of the hierarchical
scenario of cluster formation which implies that virtually every large cluster
is formed through a series of mergers of smaller progenitors. Each of these
merging events triggered violent shock waves that accelerated CR protons
through diffusive shock acceleration.  Over its cosmic history, these CRs
accumulated within the cluster volume due to their cooling time being longer
than the Hubble time \citep{1996SSRv...75..279V, 1997ApJ...487..529B}.  The
secondary radio emission probes the CR proton pressure which traces the time
integrated non-equilibrium activities of clusters and is only modulated by the
recent dynamical activities \citepalias[see also][ for average values of the
relative CR energy in different dynamical cluster
environments]{2007MNRAS...378..385P}. In our less massive clusters, the larger
scatter of the secondary emission level is due to the larger variation of
merging histories of these clusters and their weaker gravitational potential.
This leads to a larger modulation of the CR pressure and reflects more
sensitively the current merging activity of the cluster than it is the case in
large systems.  (3) In contrast to the secondary emission, the pressure of
primary CR electrons sensitively resembles the current dynamical,
non-equilibrium activity of forming structure and results in an inhomogeneous
and aspherical spatial distribution with respect to collapsed objects. This
leads to a large cluster-to-cluster variation of the primary radio emission.

{\bf Normalisation:} (1) The normalisation of the non-thermal scaling relations
depends only weakly on whether radiative or non-radiative gas physics is
simulated provided we consider in both cases only CRs from structure formation
shocks.  As discussed in \citetalias{2007PfrommerII}, this is mainly due to
self-regulated effects of the CR pressure.  The CR cooling timescales due to
Coulomb and hadronic interactions of CRs, $\tau_\rmn{pp/Coul}\propto
n_\rmn{gas}^{-1}$, adjust to different density levels in our simulations with
radiative or non-radiative gas physics.  Given a similar CR injection, this
implies a higher CR number density for a smaller gas density $n_\CR \propto
n_\rmn{gas}^{-1}$.  The secondary CR emissivities (synchrotron, IC, or pion
decay) scale as $j_\rmn{sec} \propto n_\CR n_\rmn{gas} \propto \rmn{const}$ and
remain almost invariant with respect to different gas densities.  (2) In
contrast, the normalisation sensitively depends on the assumptions and
parametrisation of the magnetic field. This clearly shows the need to
understand observationally how the properties of large scale cluster magnetic
fields vary with cluster mass and dynamical state.

{\bf Slope:} (1) The slope of the radio synchrotron scaling relations for our
{\em merging cluster sample} is largely independent of the simulated physics
or the parameters of our magnetic field if we only consider CRs from structure
formation shocks (models S1 and S2). The scaling relation is close to
$L_\rmn{NT} \propto M_\rmn{vir}^{1.5}$ (details can be found in
Tab.~\ref{tab:scaling}). The slope decreases to $\beta_\nu =1.3$ if we
additionally account for CRs from SNe feedback within galaxies. (2) If we
consider {\em all radio emitting clusters}, i.e. we also account for
radio-mini halos, the slope flattens in our radiative simulations by
$\Delta\beta_\nu \simeq 0.2$.  As a caveat, our scaling relations assume the
same parametrisation of the magnetic field for all clusters. If the central
magnetic field scales with the cluster mass, the slopes will be accordingly
steeper. Additionally, this self-similarity could be broken in the radio
synchrotron scaling relations, once magnetic field are dynamically simulated
and respond to the dynamical state of a cluster.

{\bf Scatter:} In our non-radiative simulations, the scatter in the radio
synchrotron scaling relations is much smaller than in our radiative ones.
There are no CC clusters in our non-radiative simulations by definition.  If a
merger takes place, there are stronger shock waves in our radiative simulations
due to the slightly cooler temperatures that imply smaller sound velocities and
larger Mach numbers.  Thus, the difference between relaxed and merging cluster
is more pronounced in our radiative simulations.

\subsubsection{Comparison to observations}
\label{sec:synchro_obs}

The observed sample of giant cluster radio halos \citep{2006MNRAS.369.1577C}
and that for cluster radio mini-halos \citep{2004A&A...417....1G} is compared
to our simulated scaling relations. 

{\bf Radio luminosity:} Generally, our simulated giant cluster radio halos
show the same level of radio synchrotron emission as observed ones given a
model of the magnetic field that is supported by Faraday rotation observations
\citep[][ and references therein]{2002ARA&A..40..319C, 2002RvMP...74..775W,
  2004IJMPD..13.1549G}.

{\bf Cluster magnetic fields:} The radio synchrotron emissivity scales as
\begin{equation}
j_\nu \propto \eps_\CRe\,
\eps_B^{(\alpha_\nu+1)/2} \nu^{-\alpha_\nu},
\end{equation}
where $\eps_\CRe$ and $\eps_B$ denote the energy densities of CR electrons and
magnetic fields, respectively, and the synchrotron spectral index $\alpha_\nu =
\alpha_\inj/2 = (\alpha_\e-1)/2$ is related to the spectral index of the
injected electron population $\alpha_\inj$ as well as to that of the cooled
electron population $\alpha_\e$. Typical synchrotron spectral indices of
cluster halos and relics span a range of $\alpha_\nu = 1\ldots 1.3$. This
implies a similar contribution to the radio luminosity-mass scaling relation of
clusters from the energy density of CR electrons and that of magnetic fields.
Our radio synchrotron scaling relations assume the same physical model for the
magnetic field irrespective of cluster mass and dynamical state. Conversely, we
can interpret our simulated synchrotron scaling relations as tracks in the
radio luminosity-cluster mass plane which are labelled with a set of parameters
of our magnetic model such as central magnetic field and magnetic
decline. Radio mini-halos tend to have a higher radio luminosity compared to
the giant radio halos. This hints towards a larger central magnetic field of
the order of $10\,\umu$G in these relaxed cool core clusters compared to the
apparently preferred weaker central field strength of the order of $3\,\umu$G
in merging systems. Interestingly, this characteristics of cluster magnetic
fields is also consistent with Faraday rotation measurements \citep[][ and
  references therein]{2005A&A...434...67V}.  Radio mini halos have been rarely
observed in relaxed cool core clusters. This might be partly due to the strong
radio emitting AGN at the centres of cool core clusters which implies a large
dynamic flux range to the underlying diffuse radio mini-halo and makes them
very challenging to observe.

{\bf Correlation between radio halos and mergers:} How do our simulations
support the observed radio halo-merger correlation?  (1) The radio emission
from primary, shock-accelerated electrons can boost the total radio emission
of a major merging clusters by a factor of four (cf.{\ }our massive
post-merging cluster g72a). This factor sensitively depends on the mass ratio,
geometry, and the advanced state of the merger. (2) In cool core as well as in
merging clusters, the central radio emission is dominated by synchrotron
emission from hadronically generated electrons.  In
\citetalias{2007PfrommerII}, we show that the emission size of the simulated
giant radio halos is increased due to the complex network of virializing shock
waves in the cluster periphery that are able to efficiently accelerate CR
electrons and amplify the magnetic fields due to strong shear motions.  (3)
The relative CR pressure is modulated by current merger activity of a cluster.
While this positive modulation is small in massive systems, it can be
substantial for less massive systems as can be inferred from
Fig.~\ref{fig:scalings_syn}. This is due to larger variation of merging
histories and the smaller gravitational potential in small clusters that
causes the radio emission to respond more sensitively to merging activity.

{\bf Observed scatter:} The merger causes clusters that are hosting a giant
radio halo to depart from hydrostatic equilibrium and leads to a complicated
morphology that in general is not spherical. As a result, the masses in
merging clusters can be either over- or underestimated, depending on the
amount of turbulent pressure support, the presence of shocks, and the amount
of substructure which tends to flatten the average density profile
\citep{1996ApJ...469..494E, 1996ApJ...473..651R, 2002ASSL..272..229S}.  The
sample of observed giant radio halos of \citet{2006MNRAS.369.1577C} scatters
by two orders of magnitude in synchrotron luminosity while the virial masses
of the hosting clusters only spans a factor of three.  Barring observational
uncertainties, the large range of dynamical states and merger geometries among
clusters as well as the variation of the magnetic properties such as central
field strength and magnetic decline furthermore contribute to the scatter in
the scaling relations.  The small sample size in combination with the
mentioned uncertainties make it impossible to determine a reliable
observational synchrotron scaling relation for radio halos.  Thus, the
simulated scaling relations can only be compared to the total luminosity of
the observed clusters. Studies of radio synchrotron emission from clusters
should be complemented by studies of the pixel-to-pixel correlation of
the synchrotron and X-ray surface brightness \citep{2001A&A...369..441G,
  2007PfrommerII}.


\subsection{Inverse Compton and pion decay induced $\bgamma$-ray emission}
\label{sec:gamma-rays}

In contrast to the observed diffuse radio synchrotron emission from clusters,
$\gamma$-rays from clusters have not been detected yet
\citep{2003ApJ...588..155R}. In principle, inverse Compton and pion decay
induced $\gamma$-ray emission are the cleanest way of probing structure
formation shock waves and the accelerated CR electron and proton populations
since these non-thermal emission processes are not weighted with the magnetic
energy density as it is the case for synchrotron emission.

\subsubsection{Inverse Compton and $\gamma$-ray cluster scaling relations}
\label{sec:scaling_relations}

\begin{figure*}
\begin{center}
  \begin{minipage}[t]{0.495\textwidth}
    \centering{\it \large IC emission ($E_\gamma > 10$~keV):}
  \end{minipage}
  \hfill
  \begin{minipage}[t]{0.495\textwidth}
    \centering{\it \large $\gamma$-ray emission ($E_\gamma > 100$~MeV):}
  \end{minipage}
\resizebox{0.5\hsize}{!}
{\includegraphics{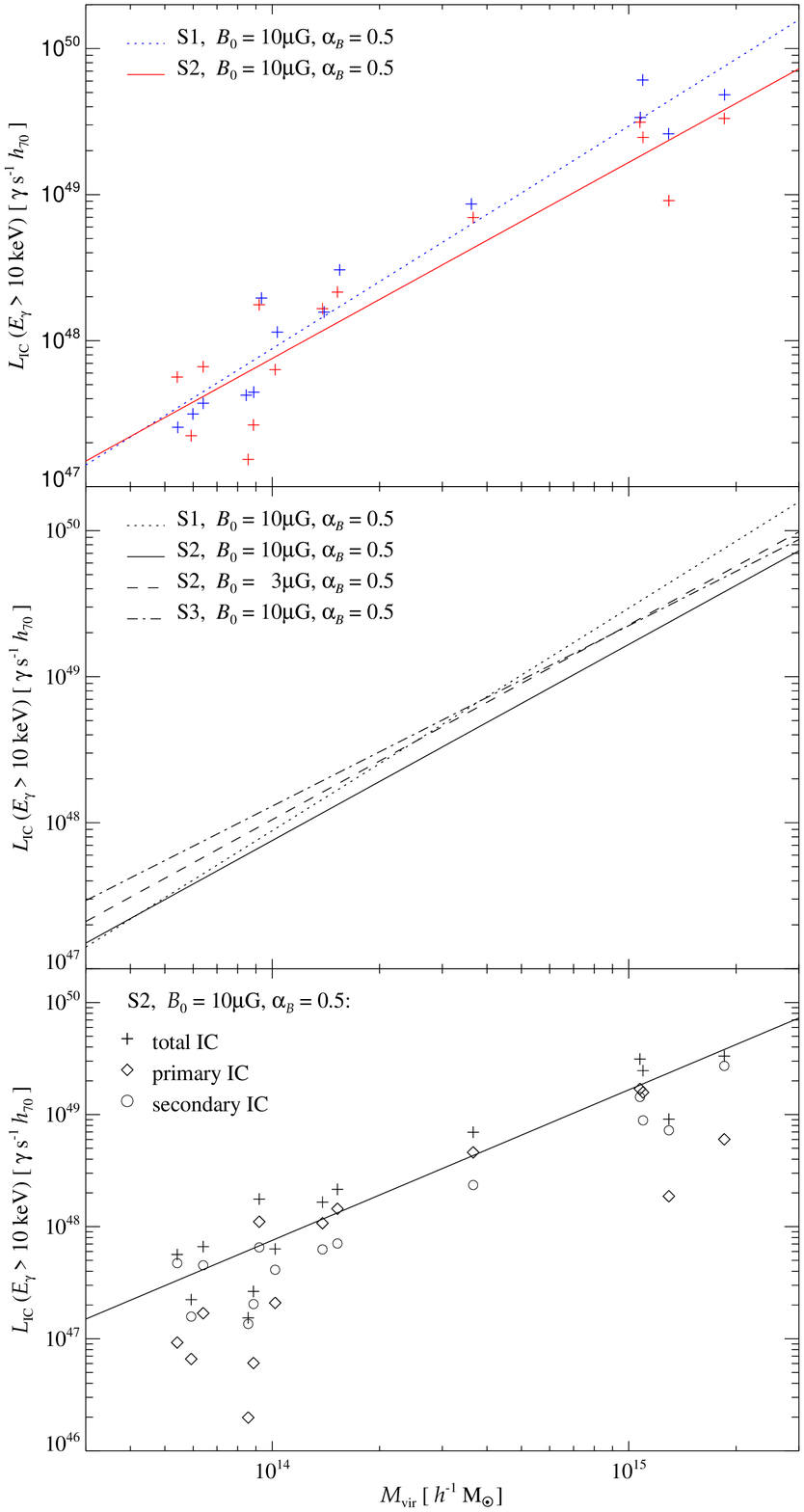}}%
\resizebox{0.5\hsize}{!}
{\includegraphics{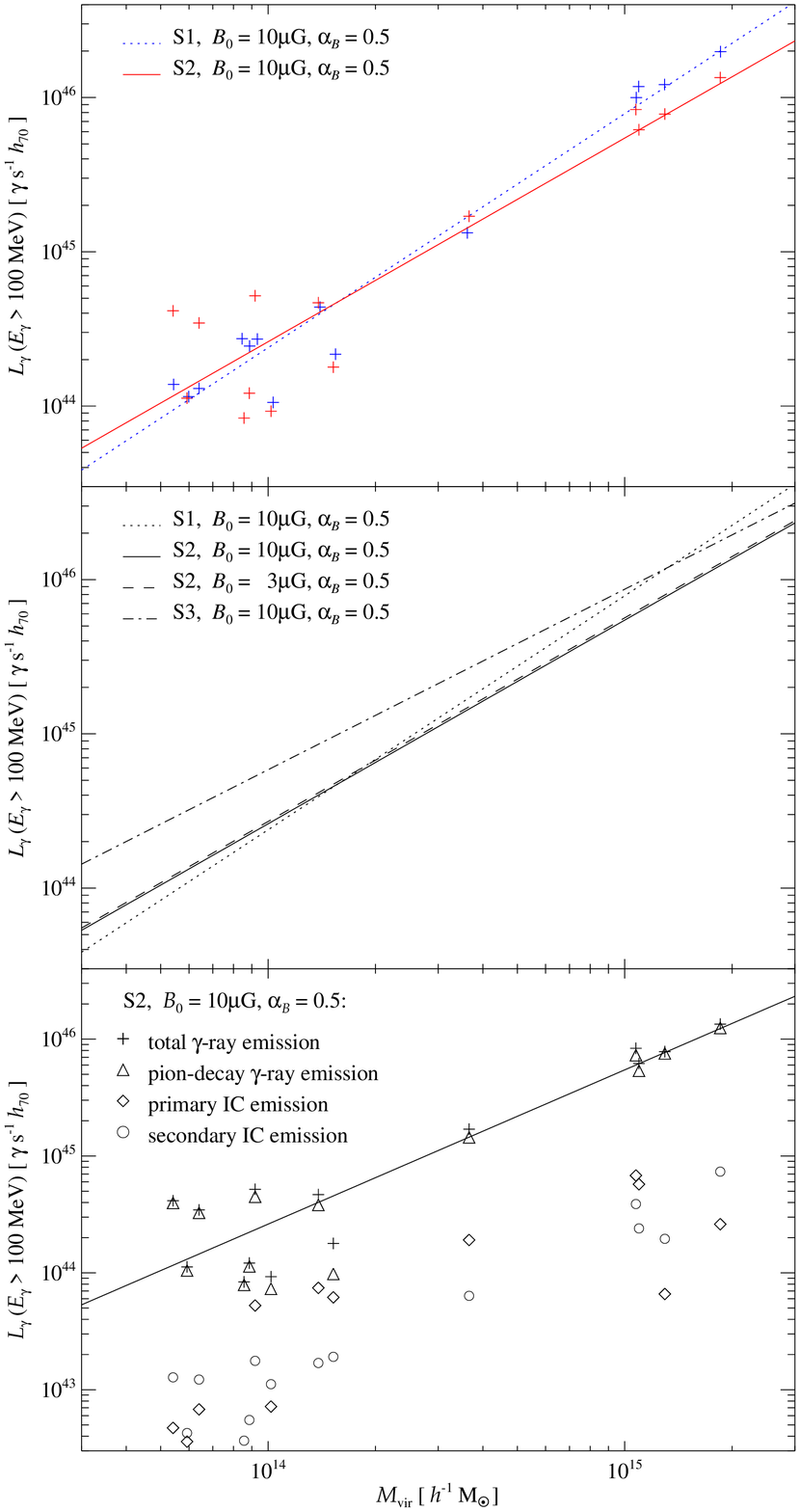}}\\
\end{center}
  \caption{Cluster scaling relations for non-thermal observables. Shown are the
    relations for the inverse Compton (IC) hard X-ray luminosities for
    $E_\gamma > 10$~keV (left panel) and the relations for the
    $\gamma$-ray luminosities for $E_\gamma > 100$~MeV (right panel). The
    top panels show the scatter of the individual clusters for our
    non-radiative (model S1) and radiative simulations (model S2). The middle
    panels show the dependence of the scaling relations on the uncertainty in
    the magnetic field and simulated physics. The bottom panels show the
    contribution of the individual emission components (primary, secondary,
    pion decay $\gamma$-rays) to the total cluster luminosities in our model S2
    while assuming a central magnetic field strength of $B_0 = 10\,\umu$G and
    an energy density scaling of $\alpha_B=0.5$.  }
  \label{fig:scalings}
\end{figure*}

We determine the cluster scaling relations for the non-thermal $\gamma$-ray
luminosities as before in Sect.~\ref{sec:radio_scaling_relations}.
Figure~\ref{fig:scalings} shows our simulated IC and pion decay scaling
relations.  The fit parameters for our models with varying simulated physics
and magnetic parameters can be found in Table~\ref{tab:scaling}.

{\bf Contributions of different emission components:} In the {\em IC scaling
relations} ($E_\rmn{IC} > 10$~keV), we see a similar picture as we found for
the radio synchrotron scaling relations, albeit somewhat amplified since the
weighting with the magnetic energy density is negligible at these energy bands.
The secondary emission component is dominant for relaxing CC clusters, and
those, that only experience a minor merger. The primary component exceeds the
secondary one for major merging clusters.  In the {\em $\gamma$-ray scaling
relations} ($E_\gamma > 100$~MeV), the pion decay component is always dominant
over the primary and secondary IC emission components. This finding does only
weakly depend on the assumed spectral index for the CR proton distribution
function since the energy band $E_\gamma > 100$~MeV is dominated by the peak of
the pion bump that is produced by GeV-protons \citep{2004A&A...413...17P}.

{\bf Normalisation:} (1) The normalisation of the non-thermal scaling relations
depends only weakly on whether radiative or non-radiative gas physics is
simulated provided we consider in both cases only CRs from structure formation
shocks. As previously discussed (cf.~Sect.~\ref{sec:radio_scaling_relations}),
this is mainly due to self-regulated effects of the CR pressure due to CR
cooling mechanisms. (2) If we additionally account for CRs from SNe feedback
within galaxies, the normalisation increases due to the second source of CR
injection. This increase is higher for our $\gamma$-ray scaling relations which
are completely dominated by the pion decay emission component.  To which extend
CRs are able to diffuse out of the cold ISM and enrich the ICM needs to be
studied separately.

{\bf Slope:} The slope of the non-thermal IC/$\gamma$-ray scaling relations
depends weakly on the simulated physics and is almost independent of the
parameters of our magnetic field. For all three non-thermal emission
mechanisms (synchrotron, IC, pion decay induced $\gamma$-ray emission), very
similar slopes are found. This is a non-trivial finding, since the relative
contribution of the various emission components differs for the different
energy bands considered in this paper.  Our set of non-radiative simulations
(S1) yields a slope of $\beta_{\rmn{IC},\gamma} \simeq 1.5$. This reduced in
our radiative simulations (S2) to $\beta_{\rmn{IC},\gamma} \simeq 1.33$ and
furthermore decreased when considering CRs from SNe feedback (S3) to
$\beta_{\rmn{IC},\gamma} \simeq 1.2$.

{\bf Scatter:} In our non-radiative simulations, the scatter in the
$\gamma$-ray scaling relations is somewhat smaller than in our radiative ones
while it is similar in the IC scaling relations. There are two reasons for
this. (1) In our non-radiative simulations, there are no CC clusters by
definition.  In merging clusters, there are stronger shock waves in our
radiative simulations due to the slightly cooler temperatures that imply
smaller sound velocities and larger Mach numbers. This leads to more effective
diffusive shock acceleration and an enhanced level of non-thermal emission.
(2) The primary emission component has its largest impact for the IC hard
X-ray emission (compared to the $\gamma$-ray emission). This component is
largely responsible for the large scatter since it traces the current
dynamical, non-equilibrium activity of the cluster.  Looking at the individual
non-thermal luminosities of our clusters (top panels in
Fig.~\ref{fig:scalings}), one can notice a large scatter. In particular for
the $\gamma$-ray emission, this scatter increases for less massive clusters in
our radiative models and can boost the $\gamma$-ray luminosity up to a factor
of four.  Due to the small sample size of our simulated high-resolution
clusters, we are unable to statistically quantify this effect reliably.

\subsubsection{Luminosity and flux functions}

\begin{figure*}
\begin{center}
  \begin{minipage}[t]{0.495\textwidth}
    \centering{\it \large $\gamma$-ray flux function ($E_\gamma > 100$~MeV):}
  \end{minipage}
  \hfill
  \begin{minipage}[t]{0.495\textwidth}
    \centering{\it \large $\gamma$-ray luminosity function ($E_\gamma > 100$~MeV):}
  \end{minipage}
\resizebox{0.5\hsize}{!}
{\includegraphics{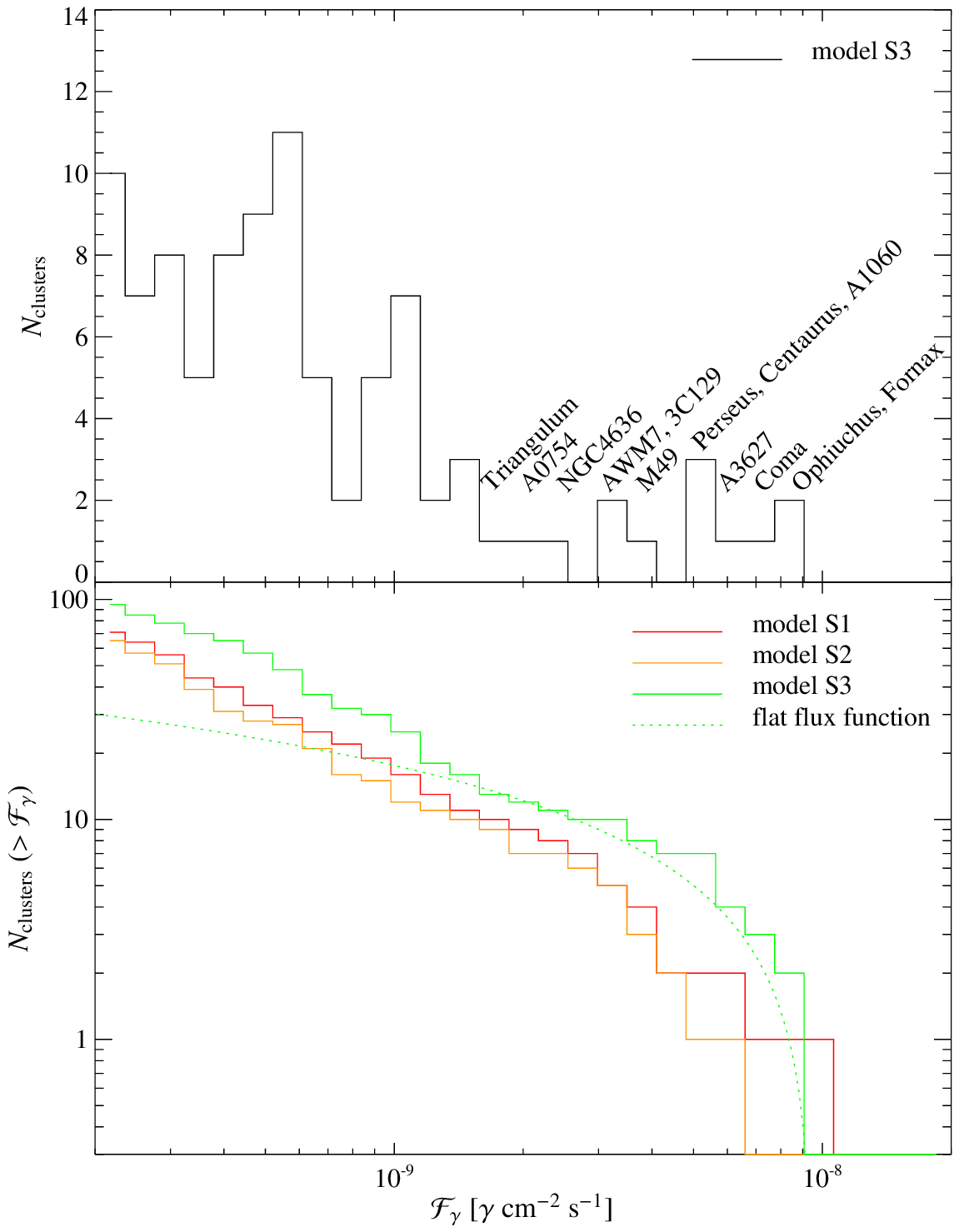}}%
\resizebox{0.5\hsize}{!}
{\includegraphics{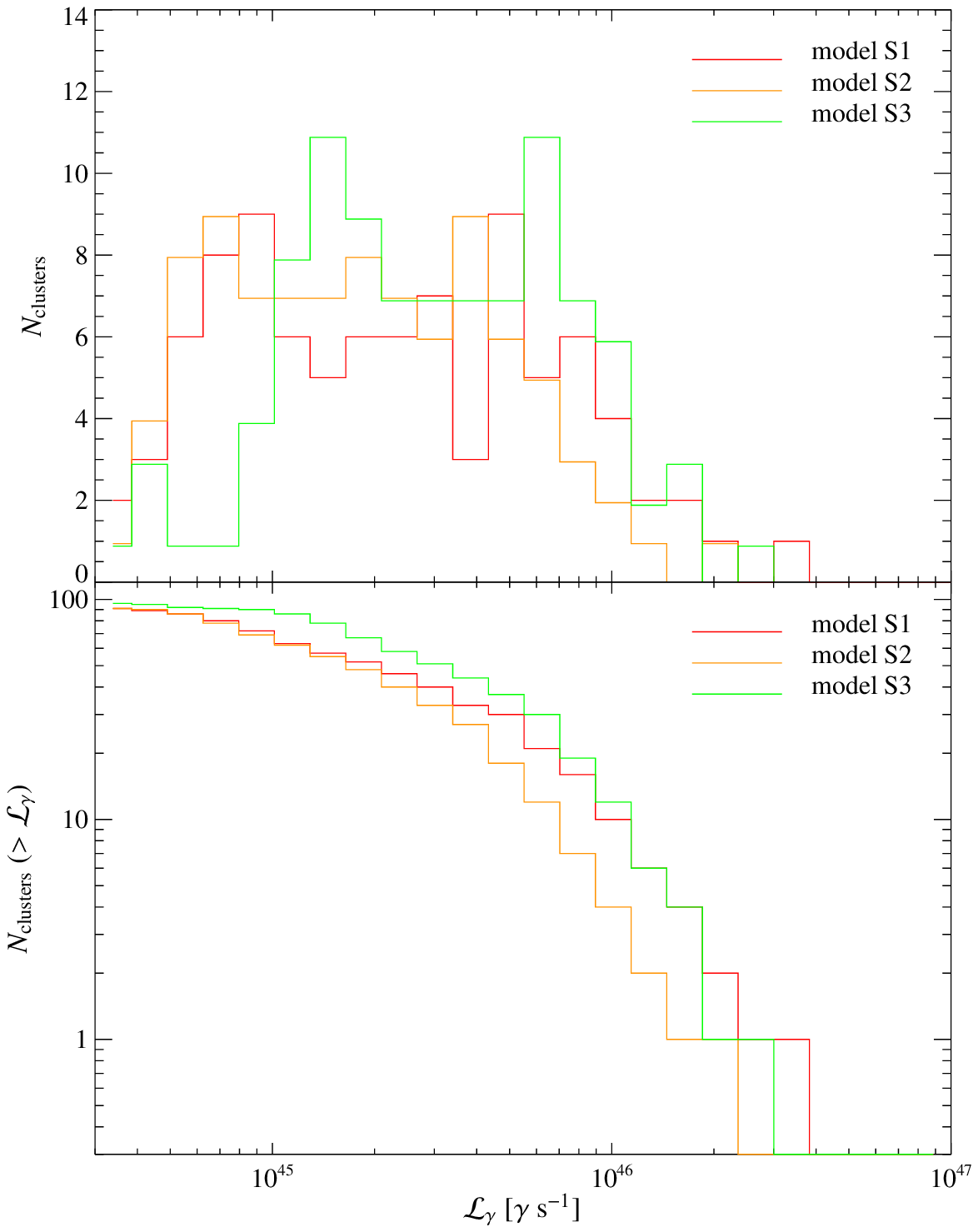}}\\
\end{center}
\caption{We use the complete sample of the X-ray brightest clusters (HIFLUGCS,
  \citet{2002ApJ...567..716R}) to predict flux and luminosity functions of the
  $\gamma$-ray emission for $E_\gamma > 100$~MeV.  The definition for our
  different models can be found in Table~\ref{tab:models}. The top panels shows
  the differential flux/luminosity functions while the bottom panels show the
  respective cumulative functions. Assuming a GLAST sensitivity after two
    years of $2\times 10^{-9} \gamma \mbox{ cm}^{-2} \mbox{ s}^{-1}$, we
  predict the detection of seven to eleven clusters named in the top left
  panel, depending on the adopted model.  }
  \label{fig:luminosity_function}
\end{figure*}

\begin{figure*}
\begin{center}
  \begin{minipage}[t]{0.495\textwidth}
    \centering{\it \large IC hard X-ray flux function ($E_\gamma > 10$~keV):}
  \end{minipage}
  \hfill
  \begin{minipage}[t]{0.495\textwidth}
    \centering{\it \large IC luminosity function ($E_\gamma > 10$~keV):}
  \end{minipage}
\resizebox{0.5\hsize}{!}
{\includegraphics{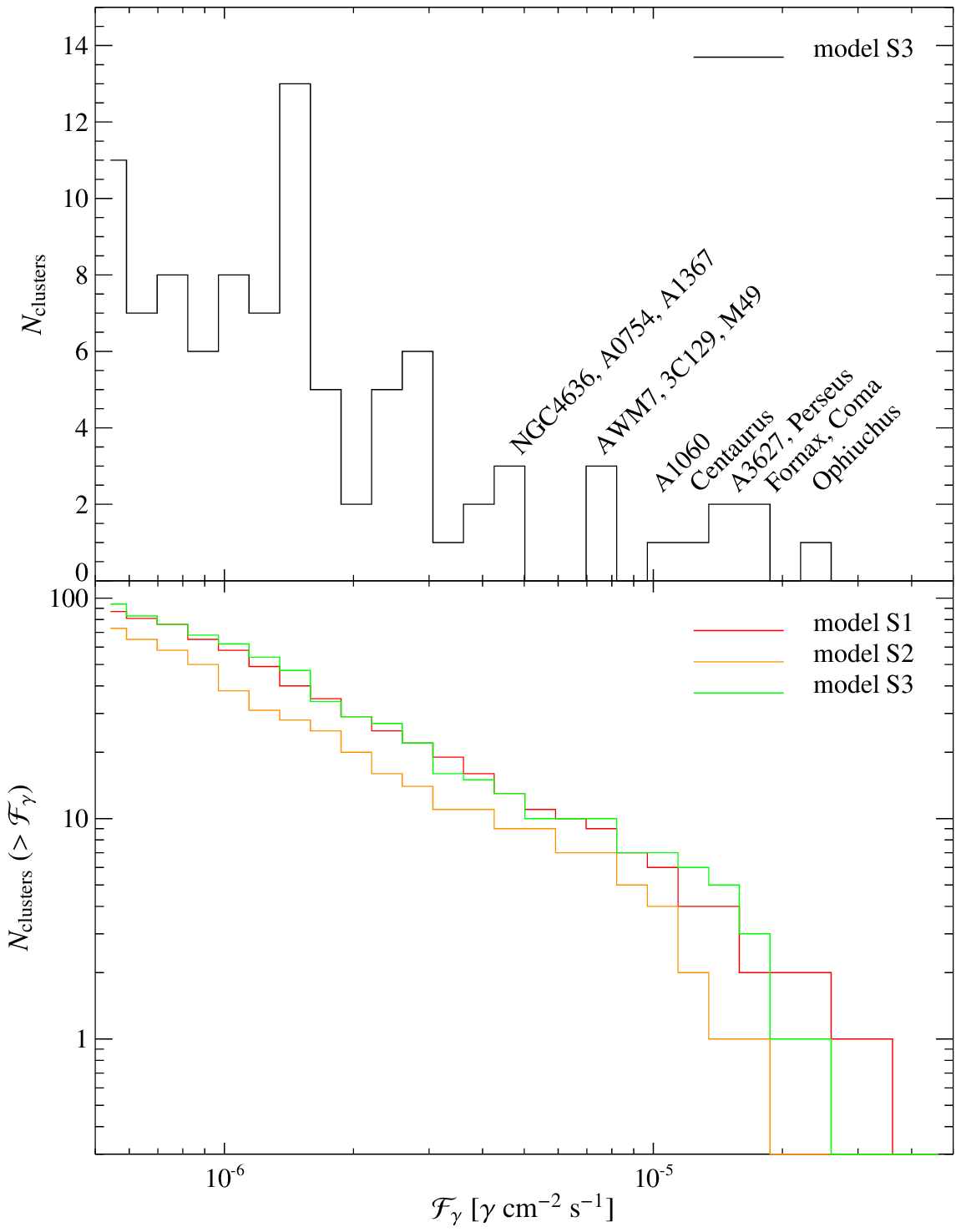}}%
\resizebox{0.5\hsize}{!}
{\includegraphics{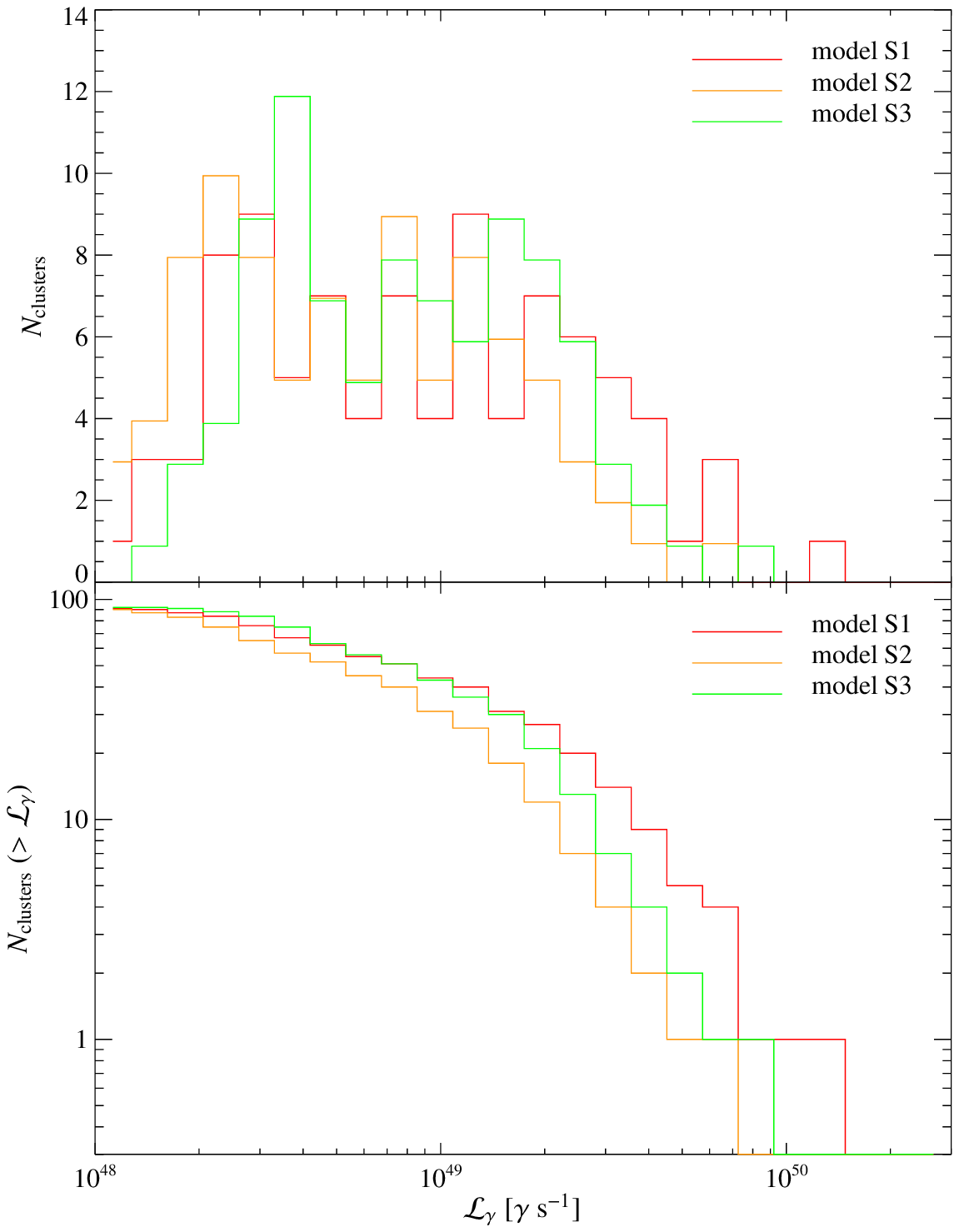}}\\
\end{center}
  \caption{We use the complete sample of the X-ray brightest clusters
    (HIFLUGCS, \citet{2002ApJ...567..716R}) to predict flux and luminosity
    functions of the hard X-ray IC emission for $E_\gamma > 10$~keV.  The
    definition for our different models can be found in
    Table~\ref{tab:models}. The top panels shows the differential
    flux/luminosity functions while the bottom panels show the respective
    cumulative functions. }
  \label{fig:IC_luminosity_function}
\end{figure*}

We combine our derived cluster scaling relations for non-thermal observables
with the complete sample of the X-ray brightest galaxy clusters (HIFLUGCS,
\citet{2002ApJ...567..716R}) to predict fluxes and luminosities of each of the
clusters. For the luminosity distance and the cluster masses, we assumed a
$\Lambda$CDM cosmology with a currently favoured Hubble constant, $h_{70}$,
where $H_0 = 70\, h_{70}\mbox{ km s}^{-1} \mbox{ Mpc}^{-1}$.  Dependent on the
simulated physics, we can thus derive flux and luminosity functions for the
$\gamma$-ray emission (Fig.~\ref{fig:luminosity_function}) and for the hard
X-ray IC emission (Fig.~\ref{fig:IC_luminosity_function}).
Tables~\ref{tab:flux} and \ref{tab:luminosity} in the Appendix show the 15
brightest as well as the 15 most luminous clusters of our homogeneous flux
limited sample.

{\bf $\bgamma$-ray emission:} Assuming a GLAST sensitivity after two years
of $2\times 10^{-9} \gamma \mbox{ cm}^{-2} \mbox{ s}^{-1}$, we predict the
detection of seven to eleven clusters named in the top left panel of
Fig.~\ref{fig:luminosity_function}, depending on the adopted model. The
brightest $\gamma$-ray clusters are Ophiuchus, Fornax, Coma, A3627, Perseus,
and Centaurus (A3526), independent of the simulated physics. Among these, only
Ophiuchus is among the ten most massive and thus most luminous clusters of the
HIFLUGCS sample.  This statement can be modified once we consider scatter in
$\gamma$-ray luminosity due the the varying dynamical states of these clusters
which might modify the rank ordering of the individual systems.

The distribution of the number of clusters with a given $\gamma$-ray flux
$\F_\gamma$ is flat in the variable $\log \F_\gamma$ down to $\F_\gamma \simeq
10^{-9} \gamma \mbox{ cm}^{-2} \mbox{ s}^{-1}$ where the true number of
clusters is suddenly increasing towards fainter fluxes. The large intrinsic
scatter around the scaling relation, especially at low $\gamma$-ray
luminosities, is expected to increase the number of cluster detections for
GLAST by scattering clusters above the survey flux limit for our case of a
decreasing differential distribution with increasing flux.\footnote{This effect
  is also known as Eddington bias \citep{1913MNRAS..73..359E}.} For comparison,
we show how a flat differential distribution with a maximum flux $\F_\rmn{max}$
translates into a cumulative one (dotted line):
\begin{equation}
  \label{eq:selection_bias}
  N(>\F) = N_0 \int_{\log \F}^\infty \theta(\log\F_\rmn{max}-x)\, \dd x = 
  N_0\,\log \frac{\F_\rmn{max}}{\F},
\end{equation}
where $\theta(x)$ denotes the Heaviside function. Any cumulative flux function
steeper than the dotted line benefits from the scatter around the scaling
relation. This has to be taken into account when deriving the observed
$\gamma$-ray luminosity function. However, due to the limited statistics in our
simulated sample, further work is needed to quantify this expected scatter.

The luminosity function shows an exponential cutoff at high $\gamma$-ray
luminosities that is inherited from the Press-Schechter mass function.  The
uncertainty at the high-mass end of our scaling relations of our different CR
models translates into a similar uncertainty of the exponential cut-off of the
cumulative luminosity function. The decrease of the luminosity function at
smaller luminosities is due to the incompleteness of the X-ray flux limited
cluster sample.

{\bf Inverse Compton emission:} We predict Ophiuchus to be the brightest hard
X-ray emitting cluster with a photon number flux of $(1.7\ldots 3.4)\times
10^{-5} \gamma \mbox{ cm}^{-2} \mbox{ s}^{-1}$ for energies $E_\gamma >
10$~keV.  Despite the fact that the derived slope $\beta_\rmn{IC}$ of the IC
scaling relation of model S3 is different compared to the $\gamma$-ray case,
all our brightest $\gamma$-ray clusters remain the brightest hard X-ray
emitting clusters.  The slopes in our other models are identical between the
$\gamma$-ray and IC case which leads to identical rank ordering of the IC
emitting clusters. Our findings with respect to the selection bias of the
number of detected clusters due to the scattering in the scaling relations
applies also in this case.

\subsubsection{Comparison to observations and previous work}

{\bf $\bgamma$-ray emission:} \citet{2003ApJ...588..155R} derived the EGRET
upper limits on the high-energy $\gamma$-ray emission of galaxy clusters using
nine years of successive observations. Stacking a sample of 58 clusters and
carefully accounting for the diffuse $\gamma$-ray background yielded an upper
2$\sigma$ limit for the average cluster of $6\times 10^{-9} \gamma \mbox{
  cm}^{-2} \mbox{ s}^{-1}$ for $E_\gamma > 100$~MeV. The limits on individual
clusters that this work predicts to be the brightest $\gamma$-ray emitters range
within $(3\ldots5)\times 10^{-8} \gamma \mbox{ cm}^{-2} \mbox{ s}^{-1}$.  Our
predicted fluxes are consistent with these upper limits, providing an
important consistency check of our models. 

\noindent
{\bf Inverse Compton emission:} There seems to be growing evidence for an
excess of hard X-ray emission compared to the expected thermal bremsstrahlung
in a number of clusters that is based on observations with instruments on board
five different X-ray satellites. Prominent examples include the Coma cluster
(\citealt{1999ApJ...511L..21R, 1999ApJ...513L..21F, 2002ApJ...579..587R,
  2004ApJ...602L..73F, 2007ApJ...654L...9F};\footnote{The results of these
  papers have been challenged by an analysis that takes into account all
  systematic uncertainties in the critical parameters including the choice of a
  source-free background field and the modelling of the thermal model for the
  ICM \citep{2004A&A...414L..41R, 2007astro.ph..2417R}.}
\citealt{2007A&A...470..835E}; using the {\em Rossi X-ray Timing Explorer}
({\em RXTE}), {\em BeppoSAX}, and {\em INTEGRAL}) and the Perseus cluster
\citep[][ using {\em Chandra} and {\em XMM-Newton}]{2005MNRAS.360..133S,
  2007Molendi}. Using our simulations, we can test the currently favoured
hypothesis that this emission is due to inverse Compton radiation by CR
electrons.  \citet{1999ApJ...513L..21F} claimed an excess flux of $2\times
10^{-11} \mbox{ erg cm}^{-2} \mbox{ s}^{-1}$ between $E_1 = 20$~keV and $E_1 =
80$~keV.  For the Coma cluster, our models predict a inverse Compton number
flux of $(1.3\ldots 2.3)\times 10^{-5} \gamma \mbox{ cm}^{-2} \mbox{ s}^{-1}$
for energies above $E_{\IC,0} = 10$~keV. To relate the number flux to an energy
flux, we assume a photon index of $\alpha_\nu=1.15$ and a scaling of $\F_\IC =
\F_0\, (E/E_{\IC,0})^{\alpha_\nu}$. Using the notation for energy and number
fluxes described in \citetalias{2007PfrommerII}, we can calculate the energy
flux in the observational hard X-ray band,
\begin{eqnarray}
  \label{eq:IC_flux}
  F_\rmn{IC} &=& \frac{\alpha_\nu}{\alpha_\nu-1}\,E_{\IC,0}\,\F_0\,
  \left[\left(\frac{E_1}{E_{\IC,0}}\right)^{1-\alpha_\nu} - 
        \left(\frac{E_2}{E_{\IC,0}}\right)^{1-\alpha_\nu}\right]\\
      &\simeq& 4\times 10^{-13} \mbox{ erg cm}^{-2} \mbox{ s}^{-1}.
      \nonumber
\end{eqnarray}
This is a factor of 50 below the claimed detection of hard X-ray emission. We
will discuss the implications of this discrepancy below.

The same argument applies to the hard X-ray emission in the Perseus cluster
where \citet{2005MNRAS.360..133S} find a flux of $6.3\times 10^{-11} \mbox{ erg
  cm}^{-2} \mbox{ s}^{-1}$ between 2 and 10~keV. Assuming a photon index of
$\alpha_\nu=1.15$, this flux exceeds our IC prediction of $5\times 10^{-13}
\mbox{ erg cm}^{-2} \mbox{ s}^{-1}$ for the same energy range by over two
orders of magnitudes. We note that in the particular case of Perseus, the main
cluster temperature of $k T_\e = 7$~keV \citep{2003ApJ...590..225C} is very
close to the energy limit of Chandra's imaging spectrometer, leaving a small
lever arm for the detection of power-law component on top of the expected
thermal bremsstrahlung components.  Assuming that the hard X-ray emission is
due to IC emission of CR electrons, we expect the non-thermal emission to be
physically and spatially unrelated to the thermal emission components.  The
morphological similarity of the high energy/temperature emission maps of
\citet{2007MNRAS.381.1381S} show a clear spatial (or angular) anti-correlation
between the hottest thermal components (4, 8 keV) and the power-law component
which questions the IC interpretation of the data.  A thorough covariance
analysis of the seven different emission components in the model of
\citet{2007MNRAS.381.1381S} would be needed in order to exclude the possibility
of component coupling in their spectral deconvolution procedure that mimics a
non-present power-law component.  The IC interpretation is also challenged on
theoretical grounds since it requires the energy density of CR electrons to be
in equipartition with the thermal plasma, leaving no room for relativistic
protons that have a much longer lifetime compared to electrons.

\noindent
{\bf Previous work:} Most of the previous work that calculated the $\gamma$-ray
emission from individual clusters made very simplifying assumptions about the
amount and spatial distribution of CRs within galaxy clusters \citep[for a
  comprehensive review, cf.][]{2007astro.ph..1545B}. Based on simplified
analytical arguments such as spherical geometry, virial equilibrium, and CRs
that are diffusing from a source in the cluster centre,
\citet{1998APh.....9..227C} derive a scaling relation of the hadronically
induced $\gamma$-ray luminosity with cluster mass $\mathcal{L} \propto M^{1/3}$
that is much shallower than our relations in Table~\ref{tab:scaling}.  The
difference can be easily explained by our more realistic simulations that
self-consistently follow the relevant CR physics leading to an inhomogeneous
distribution of relativistic protons, include hydrodynamical non-equilibrium
effects and arbitrary cluster geometries, and account for realistic
cosmological merger histories.

Modelling the non-thermal emission from clusters by numerically modelling
discretised CR energy spectra on top of Eulerian grid-based cosmological
simulations, \citet{2001ApJ...559...59M, 2001ApJ...562..233M} derive various
scaling relations of non-thermal cluster emission ranging from radio
synchrotron, IC soft and hard X-rays, to $\gamma$-rays which are in part
considerably steeper than our relations in Table~\ref{tab:scaling}.  In
contrast to our approach, these models neglected the hydrodynamic pressure of
the CR component, were quite limited in their adaptive resolution capability,
and they neglected dissipative gas physics including radiative cooling, star
formation, and supernova feedback. The cluster sample was comprised of small
systems with average core temperatures of $0.3 \mbox{ keV} < kT < 3 \mbox{
  keV}$ and non-thermal luminosities have been computed within a fixed radius
that various between $1.5 R_\rmn{vir}$ and $4 R_\rmn{vir}$ for the smallest
groups where $R_{200} \simeq 300 \,h^{-1}\mbox{ kpc}$. The discrepancy of the
non-thermal scaling relations can be understood by two main effects that lead
to an overestimation of the CR pressure inside the clusters simulated by
\citet{2001ApJ...559...59M} and thus overproduced the resulting non-thermal
emission particularly in larger systems: (1) \citet{2000ApJ...542..608M}
identified shocks with Mach numbers in the range $4 \lesssim \M \lesssim 5$ as
the most important in thermalizing the plasma. In contrast,
\citet{2003ApJ...593..599R} and \citet{2006MNRAS.367..113P} found that the Mach
number distribution peaks in the range $1 \lesssim \M \lesssim 3$.  Since
diffusive shock acceleration of CRs depends sensitively on the Mach number,
this implies a more efficient CR injection in the simulations by
\citet{2001ApJ...559...59M}. (2) The grid-based cosmological simulations have
been performed in a cosmological box of side-length $50\,h^{-1}$~Mpc with a
spatial resolution of $200\,h^{-1}$~kpc, assuming an Einstein-de Sitter
cosmological model \citep{2001ApJ...559...59M}.  The lack of resolution in the
observationally accessible, dense central regions of clusters in the grid-based
approach underestimates CR cooling processes such as Coulomb and hadronic
losses.  Secondly, these simulations are unable to resolve the adiabatic
compression of a composite of CRs and thermal gas, an effect that disfavours
the CR pressure relative to the thermal pressure.

\subsection{Minimum $\bgamma$-ray flux}
\label{sec:minimum_g-flux}

For clusters that host giant radio halos with an observed luminosity $L_\nu$,
we are able to derive a minimum $\gamma$-ray flux in the hadronic model. The
non-detection of $\gamma$-ray flux below this flux level limits the
contribution of secondary radio emission to the giant radio halo independent of
the spatial distribution of CRs and thermal gas.  The idea is based on the fact
that the radio luminosity of an equilibrium distribution of CR electrons, where
injection and cooling is balanced, becomes independent of the magnetic field in
the synchrotron dominated emission regime for $\eps_B\gg \eps_\rmn{ph}$
(cf.{\ }Fig.~3 in \citetalias{2007PfrommerII}),
\begin{eqnarray}
  L_\nu &=& 
  A_\nu \int\dd V\,C_\p n_\rmn{N}\, \frac{\eps_B}{\eps_B+\eps_\rmn{ph}}
  \left(\frac{\eps_B}{\eps_{B_\rmn{c}}}\right)^{(\alpha_\nu-1)/2}\nonumber\\
  \label{eq:hadronic_radio}
  &\simeq& A_\nu \int\dd V\,C_\p n_\rmn{N}, \quad
  \mbox{ for }\eps_B\gg \eps_\rmn{ph}\mbox{ and }\alpha_\nu\sim 1, \\
  \mathcal{L}_\gamma &=& A_\gamma \int\dd V\,C_\p n_\rmn{N},
  \label{eq:hadronic_gamma}
\end{eqnarray}
where $A_\nu$ and $A_\gamma$ are constants of the hadronic interaction physics
and given in the Appendix of \citetalias{2007PfrommerII}, the volume integral
extends over the entire cluster, $C_\p\propto n_\CR$ is the normalisation of
the CR momentum distribution and proportional to the CR number density,
$n_\rmn{N}$ is the number density of target nucleons for the hadronic
interaction, $\eps_\rmn{ph} = \eps_\rmn{CMB} + \eps_\rmn{stars}$ is the energy
density of the cosmic microwave background (CMB) and the starlight photon
field, where the equivalent magnetic field strength of the energy density of
the CMB is given by $B_\rmn{CMB} = 3.24\,\umu\mbox{G}\, (1+z)^2$, and
$\eps_{B_\rmn{c}} \simeq 31\, (\nu/\mbox{GHz})\,\umu$G is a frequency dependent
characteristic magnetic field strength for synchrotron radiation.  In this
strong field limit, the volume integral of the synchrotron emission is equal to
that of the $\gamma$-ray emission resulting from pion-decay and can be
eliminated yielding
\begin{equation}
  \label{eq:minimum_flux}
  \mathcal{F}_{\gamma,\rmn{min}} = 
  \frac{\mathcal{L}_{\gamma,\rmn{min}}}{4\upi\,D_\rmn{lum}^2} = 
  \frac{A_\gamma}{A_\nu} \frac{L_{\nu,\rmn{obs}}}{4\upi\,D_\rmn{lum}^2},
\end{equation}
and $D_\rmn{lum}$ is the luminosity distance to the cluster.  Smaller magnetic
fields would require a larger energy density of CR electrons in order to
reproduce the observed synchrotron emission and thus enhance the simultaneously
produced $\gamma$-ray emission. For the sample of known giant radio halos
\citep{2006MNRAS.369.1577C}, the Coma cluster is expected to have the largest
$\gamma$-ray flux since the combination $L_{\nu,\rmn{obs}}/D_\rmn{lum}^2$ is at
least four times larger than that in other cluster that are hosting giant radio
halos. The lowest possible hadronic $\gamma$-ray flux is realised for hard CR
spectral indices, $\alpha_\p=2$, yielding
$\mathcal{F}_{\gamma,\rmn{min}}=7.5\times 10^{-11}\gamma\mbox{ cm}^{-2}\mbox{
  s}^{-1}$ in Coma.

It turns out, that this limit can be considerably tightened by requiring the
average magnetic energy density to be locally less than the thermal energy
density. For our Coma-like cluster g72a in our simulation, a central magnetic
field strength of $10\,\umu$G corresponds to a ratio of thermal-to-magnetic
pressure of 20. Since the thermal pressure decreases by two orders of magnitude
towards the virial radius, a constant magnetic energy density (as required by
the synchrotron dominated emission regime) would exceed the thermal energy
density by a factor of five. This requires knowledge of the spatial
distribution of CRs, magnetic fields and thermal gas in our Coma-like cluster
simulation,
\begin{equation}
  \label{eq:minimum_flux_sim}
  \mathcal{F}_{\gamma,\rmn{min}} =
  \frac{L_{\nu,\rmn{obs}}}{L_{\nu,\rmn{g72a}}}\,
  \frac{\mathcal{L}_{\gamma,\rmn{g72a}}}{4\upi\,D_\rmn{lum}^2},
\end{equation}
where $L_{\nu,\rmn{g72a}}$ is the central radio halo emission due to
hadronically produced CR electrons in our model S2 (CR acceleration at
structure formation shocks while allowing for all CR loss processes).  The
predicted $\gamma$-ray luminosity in this model amounts to
$\mathcal{L}_{\gamma,\rmn{g72a}} = 7.3\times 10^{45}\,\gamma\mbox{ s}^{-1}$ and
is weakly dependent on the assumed CR spectral index of $\alpha=2.3$. The very
conservative $\gamma$-ray limit assumes a central magnetic field $B_0 =
10\,\umu$G, ensures $P_\th > 2P_B$ everywhere within the virial region of the
cluster and yields $\mathcal{F}_{\gamma,\rmn{min}} = 4\times
10^{-10}\gamma\mbox{ cm}^{-2}\mbox{ s}^{-1}$. For the same $B_0$ and $P_\th >
20 P_B$ at the virial radius, we obtain $\mathcal{F}_{\gamma,\rmn{min}} =
9\times 10^{-10}\gamma\mbox{ cm}^{-2}\mbox{ s}^{-1}$. Adopting an even lower
central magnetic field $B_0\simeq 3\,\umu$G as Faraday rotation studies of the
Coma cluster indicate \citep{1990ApJ...355...29K} and requiring $P_\th > 20
P_B$ at the virial radius, we obtain $\mathcal{F}_{\gamma,\rmn{min}} = 2\times
10^{-9}\gamma\mbox{ cm}^{-2}\mbox{ s}^{-1}= \mathcal{F}_\rmn{GLAST,~2yr}$,
i.e. the GLAST all-sky survey will be able to scrutinise this scenario after
two years. We would like to close this section by noting that our simulations
predict a $\gamma$-ray flux from Coma of $\mathcal{F}_{\gamma,\rmn{min}} =
(4\ldots7)\times 10^{-9}\gamma\mbox{ cm}^{-2}\mbox{ s}^{-1}$. This in turn
would imply a central magnetic field $B_0\simeq 3\,\umu$G with a constant
average ratio of thermal-to-magnetic pressure of 200, comparing to the observed
synchrotron flux.

\section{Discussion and Conclusions}
\label{sec:discussion}

We performed high-resolution simulations of a sample of 14 galaxy clusters that
span a mass range of almost two orders of magnitude and follow self-consistent
CR physics on top of the dissipative gas physics including radiative cooling,
star formation, and supernova feedback.  The modelled CR physics in our
simulations and our on-the-fly identification scheme of the strength of
structure formation shock waves allows us to reliably compute the relativistic
electron populations at high energies. We consider relativistic electrons that
are accelerated at cosmological structure formation shocks (so-called primary
electrons) and those that are produced in hadronic interactions of cosmic rays
with ambient gas protons (hence the name secondary or hadronic electrons).

\subsection{Non-thermal scaling relations}

In this paper, we concentrate on three observationally motivated wave-bands.
(1) Radio synchrotron emission at $1.4$~GHz, (2) non-thermal hard X-ray
emission at energies $E_\gamma>10$~keV, and (3) $\gamma$-ray emission at
energies $E_\gamma>100$~MeV.  We study the contribution of the different
emission components to the total cluster luminosity in each of these bands,
derive cluster scaling relations, and study their dependence on the simulated
physics and adopted parametrisation of the magnetic field.  Our main findings
are as follows:
\begin{enumerate}
\item The secondary emission component (radio synchrotron and inverse Compton)
is dominant for relaxing cool core clusters, and those that only experience a
minor merger.  The primary component can exceed the secondary one for major
merging clusters by a factor of four. In the $\gamma$-ray scaling
relations ($E_\gamma > 100$~MeV), the pion decay component is always dominant
over the primary and secondary IC emission components.
\item The normalisation of the non-thermal scaling relations depends only
weakly on whether radiative or non-radiative gas physics is simulated provided
we consider in both cases only CRs from structure formation shocks. This is
mainly due to self-regulated effects of the CR pressure due to the density
dependent CR cooling mechanisms.  In contrast, the normalisation of the radio
synchrotron scaling relation sensitively depends on the assumptions and
parametrisation of the magnetic field. This clearly reinforces the need to
understand observationally how the properties of large scale cluster magnetic
fields vary with cluster mass and dynamical state.
\item The slope of the non-thermal scaling relations depends weakly on the
simulated physics and is almost independent of the parameters of our magnetic
field. For all three non-thermal emission mechanisms (synchrotron, IC, pion
decay induced $\gamma$-ray emission), very similar slopes are found. This is a
non-trivial finding, since the relative contribution of the various emission
components differs for the different energy bands considered in this paper.
Our set of non-radiative simulations (S1) yields a slope of
$\beta_{\rmn{IC},\gamma} \simeq 1.5$. This is reduced in our radiative
simulations (S2) to $\beta_{\rmn{IC},\gamma} \simeq 1.33$ and furthermore
decreased when considering CRs from SNe feedback (S3) to
$\beta_{\rmn{IC},\gamma} \simeq 1.2$.  The slope of the synchrotron scaling
relation steepens if we only consider merging galaxy clusters.  As a caveat for
our synchrotron scaling relations, we assume the same parametrisation of the
magnetic field for all clusters. If the central magnetic field scales with the
cluster mass, the slopes will be accordingly steeper. Additionally, this
self-similarity could be broken in the radio synchrotron scaling relations,
once magnetic field are dynamically simulated and respond to the dynamical
state of a cluster. This reinforces the need to understand the observed scaling
properties of the magnetic field in clusters before we can draw strong
conclusions about the theory underlying the cluster radio emission.
\item In our non-radiative simulations, we observe large scatter in all
non-thermal scaling relations. This is mostly driven by active merging systems
that trigger violent shock waves and thus boost the primary emission signal.
Our results hint at a larger contribution of the scatter towards less massive
systems due to their smaller gravitational potential which needs to be checked
with a larger cluster sample size.  The large scatter will have important
implications for the number of detectable $\gamma$-ray emitting clusters by
GLAST.
\end{enumerate}

\subsection{Radio synchrotron emission}

The unified model of radio halos and relics has been put forward in our
companion paper \citepalias{2007PfrommerII} and is based on studies of the
morphology, profiles, and expected polarisation of our simulated diffuse
cluster radio synchrotron emission. The derived radio luminosities of the
primary and secondary electron populations complement this picture. We are
summarising the main findings of this work in the following.
\begin{enumerate}
\item Assuming magnetic field strengths provided by Faraday rotation
  observations, we are able to successfully reproduce the observed radio
  synchrotron luminosities of giant radio halos as well as radio mini-halos in
  our simulations. 
\item Each of our radio halo scaling relations assumes one physical model for
  the magnetic field that is described by a central field strength and magnetic
  decline. We assume it to be independent of the cluster mass and dynamical
  state. In this respect, the simulated scaling relations can be understood
  as contour lines in the radio luminosity-cluster mass plane which are
  labelled with a set of parameters of our magnetic model.  Radio mini-halos
  have a higher radio luminosity on average compared to that of giant radio
  halos.  This points towards a larger central magnetic field of the order of
  $10\,\umu$G in these relaxed cool core clusters compared to the apparently
  preferred weaker central field strength of the order of $3\,\umu$G in merging
  systems. This finding is consistent with Faraday rotation measurements and
  strongly hints at an amplifying mechanism of magnetic field strengths in
  relaxed clusters such as adiabatic compression of the fields during the
  formation of the cool core or AGN feedback amplified fields as argued in
  \citet{2006A&A...453..447E}.

\item Observed giant radio halos are all associated with merging clusters. The
  merger causes these systems to depart from hydrostatic equilibrium and leads
  to a complicated non-spherical morphology. The resulting X-ray mass
  estimates are subject to large uncertainties and might be partly responsible
  for the large scatter of observed giant radio halos of scatters by two
  orders of magnitude in synchrotron luminosity while the virial masses of the
  hosting clusters only spans a factor of three. Cluster-to-cluster variations
  of the geometry, mass ratio, the advanced state of the merger, and magnetic
  field strengths contribute furthermore to the scatter in the scaling
  relations. The small sample size in combination with the mentioned
  uncertainties doom all attempts to determine a reliable observational 
  synchrotron scaling relation for radio halos.  In contrast, studies of the
  pixel-to-pixel correlation of the synchrotron and X-ray surface brightness
  enable valuable insights that are not subject to the assumption of spherical
  symmetry.
\end{enumerate}


\subsection{Inverse Compton emission} 
 
Given our reliable modelling of the synchrotron and IC emitting high-energy CR
electron populations and our convenient parametrisation of the magnetic field
that is calibrated against Faraday rotation measurements, we can successfully
reproduce the luminosity of observed giant radio halos. However, our predicted
inverse Compton flux for the Coma and Perseus cluster falls short of the
detected excess of hard X-ray emission compared to the expected thermal
bremsstrahlung by at least a factor of 50.  Lowering the magnetic field
strength will not reconcile this discrepancy, since the IC emissivity of a
steady state electron population is independent of the magnetic energy density
in the low-field regime for $B \ll B_\rmn{CMB} = 3.24\,\umu\mbox{G}$.  This
finding can be rephrased as follows.  Combining the observed diffuse radio
synchrotron and IC emission allows to eliminate the ab initio unknown energy
density of relativistic electrons and to obtain an estimate for the magnetic
field strength that typically reaches values of $\sim 0.3\,\umu$G
\citep[e.g.,][]{1998A&A...330...90E, 1999A&A...344..409E}.  There are now three
problems associated with these low field strengths that challenge the standard
inverse Compton interpretation of the hard X-ray excess emission. (1) These
field strengths are an order of magnitude smaller than those derived from
Faraday rotation measurements which translates into two orders of magnitude in
energy density. (2) The energy density of CR electrons $\eps_\CRe$ that is in
turn needed to explain the radio halo emission would thus be two orders of
magnitudes larger than what our model of the primary and secondary electron
populations predict. (3) If we increased $\eps_\CRe$ by two orders of magnitude
(due to a different injection mechanism such as AGN jets), the resulting IC
emission at $E_\gamma > 100$~MeV would challenge upper limits on the
$\gamma$-ray emission imposed by EGRET \citep{2003ApJ...588..155R}.
Furthermore, the acceleration mechanism would have to single out fermions over
protons in order not to violate the EGRET bounds by overproducing the
simultaneously produced $\gamma$-rays from pion-decay.

There have been suggestions in the literature to circumvent the first problem:
\citet{2004JKAS...37..439E} suggests that one could in principle reconcile the
observed discrepancy of magnetic field estimates, if there is a significant
difference between volume and CRe weighted averages. This would require an
inhomogeneous magnetic energy distribution, an inhomogeneous distribution of
the CR electrons, and an anti-correlation between these two. These conditions
could be produced by physical mechanisms which produce inhomogeneous or
intermittent magnetic fields and at the same time anti-correlate the CRe
density with respect to the magnetic energy density.  As a very plausible
mechanism, he suggests synchrotron cooling in inhomogeneous magnetic fields
that provides naturally the required anti-correlation.  The hadronic model in
conjunction with peripheral shock acceleration that we studied in our series of
simulations would provide a CR electron injection rate which is not correlated
with the magnetic field strength, as would be required by the above explanation
of the discrepancy of magnetic field estimates by the two methods would
work. In contrast to this, in the re-acceleration model one would expect a
strong positive correlation of CRe and magnetic field strength, since magnetic
fields are essential for the CR electron acceleration.
\citet{2001ApJ...557..560P} alleviates the difficulties with the low magnetic
field strengths in the IC model by taking into account effects of observational
selection bias and evoking non-standard assumptions of a non-isotropic pitch
angle distribution as well as spectral breaks in the energy distribution of the
relativistic electrons.

What are the model uncertainties of our simulations that might boost the energy
density of CR electrons thus circumventing the second problem?  (1) The scatter
in the IC scaling relations seems only to be able to account for another factor
of two, albeit the small sample size of our simulated high-resolution clusters
makes it impossible to statistically quantify the scatter reliably. (2)
Adopting central magnetic field strength $B\ll B_\rmn{CMB}$ will only increase
the IC emissivity that is emitted by a steady state electron population by
another factor of two compared to our low-field case of $B \simeq B_\rmn{CMB}$
\citepalias[cf.{\ }Fig.{\ }3 in][]{2007PfrommerII}.  (3) Are there any other
sources that inject CR electrons homogeneously throughout the cluster volume
and resupply them on a time scale shorter than their radiative cooling time of
$\tau\simeq 10^8$~yrs?  CR diffusion out of AGN and radio galaxies will not
reproduce the required homogeneous distribution of CR electrons in order to
explain radio halos.  Secondly, diffusion will lead to a narrow, steep profile
of the CR electron energy density with a maximum radius of $\sqrt{\bra R^2\ket}
= \sqrt{6\, \kappa\, \tau_\rmn{cool}} = 14$~kpc, assuming a large CR
diffusivity of $\kappa = 10^{29} \mbox{ cm}^2/\mbox{s}$ and a combined
IC/synchrotron cooling time of $\tau = 10^8$~yr that corresponds to IC emitting
electrons at 10 keV with a Lorentz factor of $\gamma \simeq 3\times 10^3$ and
assuming a magnetic field of $8 \,\umu$G. This is an order of magnitude smaller
than the emission radius of the Perseus radio-mini halo and is much smaller
than the emission size of giant radio halos.  The re-acceleration model might
in principle have the correct properties to explain the spatial and spectral
electron distribution.  As laid out above, it faces however severe problems in
reconciling the observed discrepancy of magnetic field estimates from Faraday
rotation measurements on the one hand and combining synchrotron and IC
measurements one the other hand.  (4) Our model of the diffusive shock
acceleration mechanism assumes a featureless power-law \citepalias[for details,
  see][]{2007PfrommerII}. Future work will be dedicated on improving this model
to incorporate more elaborate plasma physical models.  (5) In the literature,
the excess of hard X-ray emission compared to the expected thermal
bremsstrahlung in the Coma cluster is currently controversially discussed
\citep{2004ApJ...602L..73F, 2007astro.ph..2576F, 2007ApJ...654L...9F, 
  2004A&A...414L..41R, 2007astro.ph..2417R}.  Observational efforts, such as
the future hard X-ray missions {\em NuSTAR} and {\em Simbol-X}, have to be
undertaken to unambiguously detect the spectral and spatial characteristics of
the hard X-ray excess emission.

If one relaxes the requirement of explaining the hard X-ray excess with the
same population of electrons that is responsible for the radio synchrotron
emission, there are two other models that try to explain the non-thermal excess
by synchrotron radiation of ultra-relativistic (multi-TeV) electrons and
positrons.  These electrons are continuously injected throughout the entire
intra-cluster medium either by interactions of hypothetical very high-energy
$\gamma$-rays with diffuse extragalactic radiation fields
\citep{2004A&A...417..391T} or by means of pair production processes of CMB
photons in the Coulomb field of ultra high-energetic CR (UHECR) protons that
are accelerated at structure formation shocks \citep{2005ApJ...628L...9I}.
While the generation of extremely high-energy photons remains the main
challenging question for the first model, the energy requirement of a UHECR
population of $10^{63}$~erg is rather extreme. Both models, however, are not
able to reproduce the radio-halo emission that is detected in these clusters.

Possibly the most elegant explanation for the hard X-ray emission is
non-thermal bremsstrahlung of a supra-thermal electron population that is
energised by Coulomb collisions between the quasi-thermal electrons and
non-thermal protons \citep{2007arXiv0712.0187W}. Such an electron population
displays a higher bremsstrahlung radiative efficiency than a pure power law
population thus avoiding the overheating problem of the thermal plasma
\citep{2001ApJ...557..560P}. The non-thermal protons would simultaneously be
responsible for the Coma cluster's diffuse radio halo emission
\citepalias[within the unified scheme put forward in][]{2007PfrommerII}.

\subsection{High-energy $\bgamma$-ray emission}

Our predicted $\gamma$-ray fluxes of nearby galaxy clusters are consistent with
EGRET upper limits of these clusters \citep{2003ApJ...588..155R}. The brightest
$\gamma$-ray clusters are typically a factor of five smaller than the derived
upper limits, which provides an important consistency check of our models. We
note that our simulations have not been tuned to match these upper limits,
instead we modelled the CR physics to our best knowledge and calculated the
$\gamma$-ray luminosity of our simulated clusters.

We predict the detection of the pion decay induced $\gamma$-ray emission of
seven to eleven galaxy clusters by GLAST, depending on the adopted model. The
expected brightest $\gamma$-ray clusters are Ophiuchus, Fornax, Coma, A3627,
Perseus, and Centaurus (A3526), independent of the simulated physics. Due to
the increasing slope of the differential cluster flux number distribution
towards smaller $\gamma$-ray fluxes and the large scatter in the scaling
relations (especially for less massive systems), we expect the detected number
of clusters to increase somewhat since clusters are scattered above the survey
flux limit.  For clusters that host giant radio halos, we are able to derive a
minimum $\gamma$-ray flux in the hadronic model independent of the spatial
distribution of CRs and thermal gas.  The radio luminosity of an equilibrium
distribution of CR electrons, where injection due to hadronic CR interactions
and cooling is balanced, becomes independent of the magnetic field in the
synchrotron dominated emission regime. A smaller magnetic field would require a
larger energy density of CR electrons to reproduce the observed synchrotron
luminosity and thus increase the associated $\gamma$-ray flux.  In Coma, the
absolute minimum flux of $\mathcal{F}_{\gamma,\rmn{min}} = 7.5\times
10^{-11}\gamma\mbox{ cm}^{-2}\mbox{ s}^{-1}$ is well below the sensitivity of
GLAST. Assuming magnetic field strengths as derived by Faraday rotation
measurements, these limits can be considerably tightened to match the GLAST
sensitivity after two years of all-sky survey, $\mathcal{F}_\rmn{GLAST,~2yr} =
2\times 10^{-9}\gamma\mbox{ cm}^{-2}\mbox{ s}^{-1}$.  The detection of
hadronically induced $\gamma$-ray emission will enable us to determine the CR
proton pressure in clusters and unambiguously decide upon the model of cluster
radio halos.

\section*{Acknowledgements}
I am very thankful to Volker Springel and Torsten En{\ss}lin for carefully
reading the manuscript and Torsten En{\ss}lin for suggesting the analytic
method of the minimum $\gamma$-ray flux.  All computations were performed on
CITA's McKenzie cluster \citep{2003...McKenzie} which was funded by the Canada
Foundation for Innovation and the Ontario Innovation Trust.

\bibliography{bibtex/gadget}

\begin{thebibliography}{}

\bibitem[\protect\citeauthoryear{{Aharonian}, {Akhperjanian}, {Bazer-Bachi},
  {Beilicke}, {Benbow}, {Berge}, {Bernl{\"o}hr}, {Boisson}, {Bolz}, {Borrel},
  {Braun}, {Breitling} \& {et al.}}{{Aharonian}
  et~al.}{2006}]{2006Natur.439..695A}
{Aharonian} F.,  {Akhperjanian} A.~G.,  {Bazer-Bachi} A.~R.,  {Beilicke} M.,
  {Benbow} W.,  {Berge} D.,  {Bernl{\"o}hr} K.,  {Boisson} C.,  {Bolz} O.,
  {Borrel} V.,  {Braun} I.,  {Breitling} F.,    {et al.} 2006, \nat, 439, 695

\bibitem[\protect\citeauthoryear{{Bagchi}, {Durret}, {Neto} \& {Paul}}{{Bagchi}
  et~al.}{2006}]{2006Sci...314..791B}
{Bagchi} J.,  {Durret} F.,  {Neto} G.~B.~L.,    {Paul} S.,  2006, Science, 314,
  791

\bibitem[\protect\citeauthoryear{{Berezinsky}, {Blasi} \&
  {Ptuskin}}{{Berezinsky} et~al.}{1997}]{1997ApJ...487..529B}
{Berezinsky} V.~S.,  {Blasi} P.,    {Ptuskin} V.~S.,  1997, \apj, 487, 529

\bibitem[\protect\citeauthoryear{{Blasi} \& {Colafrancesco}}{{Blasi} \&
  {Colafrancesco}}{1999}]{1999APh....12..169B}
{Blasi} P.,  {Colafrancesco} S.,  1999, Astroparticle Physics, 12, 169

\bibitem[\protect\citeauthoryear{{Blasi}, {Gabici} \& {Brunetti}}{{Blasi}
  et~al.}{2007}]{2007astro.ph..1545B}
{Blasi} P.,  {Gabici} S.,    {Brunetti} G.,  2007, ArXiv Astrophysics e-prints

\bibitem[\protect\citeauthoryear{{Brunetti}, {Blasi}, {Cassano} \&
  {Gabici}}{{Brunetti} et~al.}{2004}]{2004MNRAS.350.1174B}
{Brunetti} G.,  {Blasi} P.,  {Cassano} R.,    {Gabici} S.,  2004, \mnras, 350,
  1174

\bibitem[\protect\citeauthoryear{{Brunetti} \& {Lazarian}}{{Brunetti} \&
  {Lazarian}}{2007}]{2007MNRAS.378..245B}
{Brunetti} G.,  {Lazarian} A.,  2007, \mnras, 378, 245

\bibitem[\protect\citeauthoryear{{Brunetti}, {Setti}, {Feretti} \&
  {Giovannini}}{{Brunetti} et~al.}{2001}]{2001MNRAS.320..365B}
{Brunetti} G.,  {Setti} G.,  {Feretti} L.,    {Giovannini} G.,  2001, \mnras,
  320, 365

\bibitem[\protect\citeauthoryear{{Carilli} \& {Taylor}}{{Carilli} \&
  {Taylor}}{2002}]{2002ARA&A..40..319C}
{Carilli} C.~L.,  {Taylor} G.~B.,  2002, \araa, 40, 319

\bibitem[\protect\citeauthoryear{{Cassano}, {Brunetti} \& {Setti}}{{Cassano}
  et~al.}{2006}]{2006MNRAS.369.1577C}
{Cassano} R.,  {Brunetti} G.,    {Setti} G.,  2006, \mnras, 369, 1577

\bibitem[\protect\citeauthoryear{{Churazov}, {Forman}, {Jones} \& {B{\"
  o}hringer}}{{Churazov} et~al.}{2003}]{2003ApJ...590..225C}
{Churazov} E.,  {Forman} W.,  {Jones} C.,    {B{\" o}hringer} H.,  2003, \apj,
  590, 225

\bibitem[\protect\citeauthoryear{{Clarke} \& {En{\ss}lin}}{{Clarke} \&
  {En{\ss}lin}}{2006}]{2006AJ....131.2900C}
{Clarke} T.~E.,  {En{\ss}lin} T.~A.,  2006, \aj, 131, 2900

\bibitem[\protect\citeauthoryear{{Colafrancesco} \& {Blasi}}{{Colafrancesco} \&
  {Blasi}}{1998}]{1998APh.....9..227C}
{Colafrancesco} S.,  {Blasi} P.,  1998, Astroparticle Physics, 9, 227

\bibitem[\protect\citeauthoryear{{Deiss}, {Reich}, {Lesch} \&
  {Wielebinski}}{{Deiss} et~al.}{1997}]{1997A&A...321...55D}
{Deiss} B.~M.,  {Reich} W.,  {Lesch} H.,    {Wielebinski} R.,  1997, \aap, 321,
  55

\bibitem[\protect\citeauthoryear{{Dennison}}{{Dennison}}{1980}]{1980ApJ...239L%
..93D}
{Dennison} B.,  1980, \apjl, 239, L93

\bibitem[\protect\citeauthoryear{{Dolag}, {Bartelmann} \& {Lesch}}{{Dolag}
  et~al.}{1999}]{1999A&A...348..351D}
{Dolag} K.,  {Bartelmann} M.,    {Lesch} H.,  1999, \aap, 348, 351

\bibitem[\protect\citeauthoryear{{Dolag} \& {En{\ss}lin}}{{Dolag} \&
  {En{\ss}lin}}{2000}]{2000A&A...362..151D}
{Dolag} K.,  {En{\ss}lin} T.~A.,  2000, \aap, 362, 151

\bibitem[\protect\citeauthoryear{{Dolag}, {Schindler}, {Govoni} \&
  {Feretti}}{{Dolag} et~al.}{2001}]{2001A&A...378..777D}
{Dolag} K.,  {Schindler} S.,  {Govoni} F.,    {Feretti} L.,  2001, \aap, 378,
  777

\bibitem[\protect\citeauthoryear{{Dubinski}, {Humble}, {Loken}, {Pen} \&
  {Martin}}{{Dubinski} et~al.}{2003}]{2003...McKenzie}
{Dubinski} J.,  {Humble} R.~J.,  {Loken} C.,  {Pen} U.-L.,    {Martin} P.~G.,
  2003, in Proc. of the 17th Annual International Symposium on High Performance
  Computing Systems and Applications: {McKenzie: A Teraflops Linux Beowulf
  Cluster for Computational Astrophysics}

\bibitem[\protect\citeauthoryear{{Eckert}, {Neronov}, {Courvoisier} \&
  {Produit}}{{Eckert} et~al.}{2007}]{2007A&A...470..835E}
{Eckert} D.,  {Neronov} A.,  {Courvoisier} T.~J.-L.,    {Produit} N.,  2007,
  \aap, 470, 835

\bibitem[\protect\citeauthoryear{{Eddington}}{{Eddington}}{1913}]{1913MNRAS..7%
3..359E}
{Eddington} A.~S.,  1913, \mnras, 73, 359

\bibitem[\protect\citeauthoryear{{Eke}, {Cole} \& {Frenk}}{{Eke}
  et~al.}{1996}]{1996MNRAS.282..263E}
{Eke} V.~R.,  {Cole} S.,    {Frenk} C.~S.,  1996, \mnras, 282, 263

\bibitem[\protect\citeauthoryear{{En{\ss}lin}}{{En{\ss}lin}}{2004}]{2004JKAS..%
.37..439E}
{En{\ss}lin} T.,  2004, Journal of Korean Astronomical Society, 37, 439

\bibitem[\protect\citeauthoryear{{En{\ss}lin} \& {Biermann}}{{En{\ss}lin} \&
  {Biermann}}{1998}]{1998A&A...330...90E}
{En{\ss}lin} T.~A.,  {Biermann} P.~L.,  1998, \aap, 330, 90

\bibitem[\protect\citeauthoryear{{Ensslin}, {Biermann}, {Klein} \&
  {Kohle}}{{Ensslin} et~al.}{1998}]{1998A&A...332..395E}
{Ensslin} T.~A.,  {Biermann} P.~L.,  {Klein} U.,    {Kohle} S.,  1998, \aap,
  332, 395

\bibitem[\protect\citeauthoryear{{En{\ss}lin}, {Lieu} \&
  {Biermann}}{{En{\ss}lin} et~al.}{1999}]{1999A&A...344..409E}
{En{\ss}lin} T.~A.,  {Lieu} R.,    {Biermann} P.~L.,  1999, \aap, 344, 409

\bibitem[\protect\citeauthoryear{{En{\ss}lin}, {Pfrommer}, {Springel} \&
  {Jubelgas}}{{En{\ss}lin} et~al.}{2007}]{2007A&A...473...41E}
{En{\ss}lin} T.~A.,  {Pfrommer} C.,  {Springel} V.,    {Jubelgas} M.,  2007,
  \aap, 473, 41

\bibitem[\protect\citeauthoryear{{En{\ss}lin} \& {Vogt}}{{En{\ss}lin} \&
  {Vogt}}{2006}]{2006A&A...453..447E}
{En{\ss}lin} T.~A.,  {Vogt} C.,  2006, \aap, 453, 447

\bibitem[\protect\citeauthoryear{{Evrard}, {Metzler} \& {Navarro}}{{Evrard}
  et~al.}{1996}]{1996ApJ...469..494E}
{Evrard} A.~E.,  {Metzler} C.~A.,    {Navarro} J.~F.,  1996, \apj, 469, 494

\bibitem[\protect\citeauthoryear{{Feretti}, {Brunetti}, {Giovannini}, {Kassim},
  {Orr{\'u}} \& {Setti}}{{Feretti} et~al.}{2004}]{2004JKAS...37..315F}
{Feretti} L.,  {Brunetti} G.,  {Giovannini} G.,  {Kassim} N.,  {Orr{\'u}} E.,
   {Setti} G.,  2004, Journal of Korean Astronomical Society, 37, 315

\bibitem[\protect\citeauthoryear{{Fusco-Femiano}, {dal Fiume}, {Feretti},
  {Giovannini}, {Grandi}, {Matt}, {Molendi} \& {Santangelo}}{{Fusco-Femiano}
  et~al.}{1999}]{1999ApJ...513L..21F}
{Fusco-Femiano} R.,  {dal Fiume} D.,  {Feretti} L.,  {Giovannini} G.,  {Grandi}
  P.,  {Matt} G.,  {Molendi} S.,    {Santangelo} A.,  1999, \apjl, 513, L21

\bibitem[\protect\citeauthoryear{{Fusco-Femiano}, {Landi} \&
  {Orlandini}}{{Fusco-Femiano} et~al.}{2007a}]{2007astro.ph..2576F}
{Fusco-Femiano} R.,  {Landi} R.,    {Orlandini} M.,  2007a, astro-ph/0702576

\bibitem[\protect\citeauthoryear{{Fusco-Femiano}, {Landi} \&
  {Orlandini}}{{Fusco-Femiano} et~al.}{2007b}]{2007ApJ...654L...9F}
{Fusco-Femiano} R.,  {Landi} R.,    {Orlandini} M.,  2007b, \apjl, 654, L9

\bibitem[\protect\citeauthoryear{{Fusco-Femiano}, {Orlandini}, {Brunetti},
  {Feretti}, {Giovannini}, {Grandi} \& {Setti}}{{Fusco-Femiano}
  et~al.}{2004}]{2004ApJ...602L..73F}
{Fusco-Femiano} R.,  {Orlandini} M.,  {Brunetti} G.,  {Feretti} L.,
  {Giovannini} G.,  {Grandi} P.,    {Setti} G.,  2004, \apjl, 602, L73

\bibitem[\protect\citeauthoryear{{Giovannini}, {Tordi} \&
  {Feretti}}{{Giovannini} et~al.}{1999}]{1999NewA....4..141G}
{Giovannini} G.,  {Tordi} M.,    {Feretti} L.,  1999, New Astronomy, 4, 141

\bibitem[\protect\citeauthoryear{{Gitti}, {Brunetti}, {Feretti} \&
  {Setti}}{{Gitti} et~al.}{2004}]{2004A&A...417....1G}
{Gitti} M.,  {Brunetti} G.,  {Feretti} L.,    {Setti} G.,  2004, \aap, 417, 1

\bibitem[\protect\citeauthoryear{{Gitti}, {Ferrari}, {Domainko}, {Feretti} \&
  {Schindler}}{{Gitti} et~al.}{2007}]{2007A&A...470L..25G}
{Gitti} M.,  {Ferrari} C.,  {Domainko} W.,  {Feretti} L.,    {Schindler} S.,
  2007, \aap, 470, L25

\bibitem[\protect\citeauthoryear{{Govoni}, {En{\ss}lin}, {Feretti} \&
  {Giovannini}}{{Govoni} et~al.}{2001}]{2001A&A...369..441G}
{Govoni} F.,  {En{\ss}lin} T.~A.,  {Feretti} L.,    {Giovannini} G.,  2001,
  \aap, 369, 441

\bibitem[\protect\citeauthoryear{{Govoni} \& {Feretti}}{{Govoni} \&
  {Feretti}}{2004}]{2004IJMPD..13.1549G}
{Govoni} F.,  {Feretti} L.,  2004, International Journal of Modern Physics D,
  13, 1549

\bibitem[\protect\citeauthoryear{{Inoue}, {Aharonian} \& {Sugiyama}}{{Inoue}
  et~al.}{2005}]{2005ApJ...628L...9I}
{Inoue} S.,  {Aharonian} F.~A.,    {Sugiyama} N.,  2005, \apjl, 628, L9

\bibitem[\protect\citeauthoryear{{Jaffe}}{{Jaffe}}{1977}]{1977ApJ...212....1J}
{Jaffe} W.~J.,  1977, \apj, 212, 1

\bibitem[\protect\citeauthoryear{{Jubelgas}, {Springel}, {En{\ss}lin} \&
  {Pfrommer}}{{Jubelgas} et~al.}{2007}]{2006...Jubelgas}
{Jubelgas} M.,  {Springel} V.,  {En{\ss}lin} T.~A.,    {Pfrommer} C.,  2007,
  \aap, in print, arXiv:astro-ph/0603485

\bibitem[\protect\citeauthoryear{{Kempner}, {Blanton}, {Clarke}, {En{\ss}lin},
  {Johnston-Hollitt} \& {Rudnick}}{{Kempner}
  et~al.}{2004}]{2004rcfg.proc..335K}
{Kempner} J.~C.,  {Blanton} E.~L.,  {Clarke} T.~E.,  {En{\ss}lin} T.~A.,
  {Johnston-Hollitt} M.,    {Rudnick} L.,  2004, in The Riddle of Cooling Flows
  in Galaxies and Clusters of galaxies, {Conference Note: A Taxonomy of
  Extended Radio Sources in Clusters of Galaxies}.
p.~335

\bibitem[\protect\citeauthoryear{{Kim}, {Kronberg}, {Giovannini} \&
  {Venturi}}{{Kim} et~al.}{1989}]{1989Natur.341..720K}
{Kim} K.~.,  {Kronberg} P.~P.,  {Giovannini} G.,    {Venturi} T.,  1989, \nat,
  341, 720

\bibitem[\protect\citeauthoryear{{Kim}, {Kronberg}, {Dewdney} \&
  {Landecker}}{{Kim} et~al.}{1990}]{1990ApJ...355...29K}
{Kim} K.-T.,  {Kronberg} P.~P.,  {Dewdney} P.~E.,    {Landecker} T.~L.,  1990,
  \apj, 355, 29

\bibitem[\protect\citeauthoryear{{Liang}, {Hunstead}, {Birkinshaw} \&
  {Andreani}}{{Liang} et~al.}{2000}]{2000ApJ...544..686L}
{Liang} H.,  {Hunstead} R.~W.,  {Birkinshaw} M.,    {Andreani} P.,  2000, \apj,
  544, 686

\bibitem[\protect\citeauthoryear{{Miniati}, {Ryu}, {Kang} \& {Jones}}{{Miniati}
  et~al.}{2001a}]{2001ApJ...559...59M}
{Miniati} F.,  {Ryu} D.,  {Kang} H.,    {Jones} T.~W.,  2001a, \apj, 559, 59

\bibitem[\protect\citeauthoryear{{Miniati}, {Jones}, {Kang} \& {Ryu}}{{Miniati}
  et~al.}{2001b}]{2001ApJ...562..233M}
{Miniati} F.,  {Jones} T.~W.,  {Kang} H.,    {Ryu} D.,  2001b, \apj, 562, 233

\bibitem[\protect\citeauthoryear{{Miniati}, {Ryu}, {Kang}, {Jones}, {Cen} \&
  {Ostriker}}{{Miniati} et~al.}{2000}]{2000ApJ...542..608M}
{Miniati} F.,  {Ryu} D.,  {Kang} H.,  {Jones} T.~W.,  {Cen} R.,    {Ostriker}
  J.~P.,  2000, \apj, 542, 608

\bibitem[\protect\citeauthoryear{{Molendi}}{{Molendi}}{2007}]{2007Molendi}
{Molendi} S.,  2007, private communication

\bibitem[\protect\citeauthoryear{{Ohno}, {Takizawa} \& {Shibata}}{{Ohno}
  et~al.}{2002}]{2002ApJ...577..658O}
{Ohno} H.,  {Takizawa} M.,    {Shibata} S.,  2002, \apj, 577, 658

\bibitem[\protect\citeauthoryear{{Pedlar}, {Ghataure}, {Davies}, {Harrison},
  {Perley}, {Crane} \& {Unger}}{{Pedlar} et~al.}{1990}]{1990MNRAS.246..477P}
{Pedlar} A.,  {Ghataure} H.~S.,  {Davies} R.~D.,  {Harrison} B.~A.,  {Perley}
  R.,  {Crane} P.~C.,    {Unger} S.~W.,  1990, \mnras, 246, 477

\bibitem[\protect\citeauthoryear{{Petrosian}}{{Petrosian}}{2001}]{2001ApJ...55%
7..560P}
{Petrosian} V.,  2001, \apj, 557, 560

\bibitem[\protect\citeauthoryear{{Pfrommer} \& {En{\ss}lin}}{{Pfrommer} \&
  {En{\ss}lin}}{2003}]{2003A&A...407L..73P}
{Pfrommer} C.,  {En{\ss}lin} T.~A.,  2003, \aap, 407, L73

\bibitem[\protect\citeauthoryear{{Pfrommer} \& {En{\ss}lin}}{{Pfrommer} \&
  {En{\ss}lin}}{2004a}]{2004A&A...413...17P}
{Pfrommer} C.,  {En{\ss}lin} T.~A.,  2004a, \aap, 413, 17

\bibitem[\protect\citeauthoryear{{Pfrommer} \& {En{\ss}lin}}{{Pfrommer} \&
  {En{\ss}lin}}{2004b}]{2004MNRAS.352...76P}
{Pfrommer} C.,  {En{\ss}lin} T.~A.,  2004b, \mnras, 352, 76

\bibitem[\protect\citeauthoryear{{Pfrommer}, {En{\ss}lin} \&
  {Springel}}{{Pfrommer} et~al.}{2007b}]{2007PfrommerII}
{Pfrommer} C.,  {En{\ss}lin} T.~A.,    {Springel} 2007b, ArXiv:0707.1707

\bibitem[\protect\citeauthoryear{{Pfrommer}, {En{\ss}lin}, {Springel},
  {Jubelgas} \& {Dolag}}{{Pfrommer} et~al.}{2007a}]{2007MNRAS...378..385P}
{Pfrommer} C.,  {En{\ss}lin} T.~A.,  {Springel} V.,  {Jubelgas} M.,    {Dolag}
  K.,  2007a, \mnras, 378, 385

\bibitem[\protect\citeauthoryear{{Pfrommer}, {Springel}, {En{\ss}lin} \&
  {Jubelgas}}{{Pfrommer} et~al.}{2006}]{2006MNRAS.367..113P}
{Pfrommer} C.,  {Springel} V.,  {En{\ss}lin} T.~A.,    {Jubelgas} M.,  2006,
  \mnras, 367, 113

\bibitem[\protect\citeauthoryear{{Reimer}, {Pohl}, {Sreekumar} \&
  {Mattox}}{{Reimer} et~al.}{2003}]{2003ApJ...588..155R}
{Reimer} O.,  {Pohl} M.,  {Sreekumar} P.,    {Mattox} J.~R.,  2003, \apj, 588,
  155

\bibitem[\protect\citeauthoryear{{Reiprich} \& {B{\"o}hringer}}{{Reiprich} \&
  {B{\"o}hringer}}{2002}]{2002ApJ...567..716R}
{Reiprich} T.~H.,  {B{\"o}hringer} H.,  2002, \apj, 567, 716

\bibitem[\protect\citeauthoryear{{Rephaeli} \& {Gruber}}{{Rephaeli} \&
  {Gruber}}{2002}]{2002ApJ...579..587R}
{Rephaeli} Y.,  {Gruber} D.,  2002, \apj, 579, 587

\bibitem[\protect\citeauthoryear{{Rephaeli}, {Gruber} \& {Blanco}}{{Rephaeli}
  et~al.}{1999}]{1999ApJ...511L..21R}
{Rephaeli} Y.,  {Gruber} D.,    {Blanco} P.,  1999, \apjl, 511, L21

\bibitem[\protect\citeauthoryear{{Roettiger}, {Burns} \& {Loken}}{{Roettiger}
  et~al.}{1996}]{1996ApJ...473..651R}
{Roettiger} K.,  {Burns} J.~O.,    {Loken} C.,  1996, \apj, 473, 651

\bibitem[\protect\citeauthoryear{{Rossetti} \& {Molendi}}{{Rossetti} \&
  {Molendi}}{2004}]{2004A&A...414L..41R}
{Rossetti} M.,  {Molendi} S.,  2004, \aap, 414, L41

\bibitem[\protect\citeauthoryear{{Rossetti} \& {Molendi}}{{Rossetti} \&
  {Molendi}}{2007}]{2007astro.ph..2417R}
{Rossetti} M.,  {Molendi} S.,  2007, astro-ph/0702417

\bibitem[\protect\citeauthoryear{{R{\"o}ttgering}, {Wieringa}, {Hunstead} \&
  {Ekers}}{{R{\"o}ttgering} et~al.}{1997}]{1997MNRAS.290..577R}
{R{\"o}ttgering} H.~J.~A.,  {Wieringa} M.~H.,  {Hunstead} R.~W.,    {Ekers}
  R.~D.,  1997, \mnras, 290, 577

\bibitem[\protect\citeauthoryear{{Ryu}, {Kang}, {Hallman} \& {Jones}}{{Ryu}
  et~al.}{2003}]{2003ApJ...593..599R}
{Ryu} D.,  {Kang} H.,  {Hallman} E.,    {Jones} T.~W.,  2003, \apj, 593, 599

\bibitem[\protect\citeauthoryear{{Sanders} \& {Fabian}}{{Sanders} \&
  {Fabian}}{2007}]{2007MNRAS.381.1381S}
{Sanders} J.~S.,  {Fabian} A.~C.,  2007, \mnras, 381, 1381

\bibitem[\protect\citeauthoryear{{Sanders}, {Fabian} \& {Dunn}}{{Sanders}
  et~al.}{2005}]{2005MNRAS.360..133S}
{Sanders} J.~S.,  {Fabian} A.~C.,    {Dunn} R.~J.~H.,  2005, \mnras, 360, 133

\bibitem[\protect\citeauthoryear{{Sarazin}}{{Sarazin}}{2002}]{2002mpgc.book...%
.1S}
{Sarazin} C.~L.,  2002, in ASSL Vol. 272: Merging Processes in Galaxy Clusters
  {The Physics of Cluster Mergers}.
pp 1--38

\bibitem[\protect\citeauthoryear{{Schekochihin} \& {Cowley}}{{Schekochihin} \&
  {Cowley}}{2006}]{2006PhPl...13e6501S}
{Schekochihin} A.~A.,  {Cowley} S.~C.,  2006, Physics of Plasmas, 13, 6501

\bibitem[\protect\citeauthoryear{{Schindler}}{{Schindler}}{2002}]{2002ASSL..27%
2..229S}
{Schindler} S.,  2002, in {Feretti} L.,  {Gioia} I.~M.,   {Giovannini} G.,
  eds, Astrophysics and Space Science Library Vol.~272 of Astrophysics and
  Space Science Library, {Mergers of Galaxy Clusters in Numerical Simulations}.
pp 229--251

\bibitem[\protect\citeauthoryear{{Schlickeiser}, {Sievers} \&
  {Thiemann}}{{Schlickeiser} et~al.}{1987}]{1987A&A...182...21S}
{Schlickeiser} R.,  {Sievers} A.,    {Thiemann} H.,  1987, \aap, 182, 21

\bibitem[\protect\citeauthoryear{{Subramanian}}{{Subramanian}}{2003}]{2003PhRv%
L..90x5003S}
{Subramanian} K.,  2003, Physical Review Letters, 90, 245003

\bibitem[\protect\citeauthoryear{{Timokhin}, {Aharonian} \&
  {Neronov}}{{Timokhin} et~al.}{2004}]{2004A&A...417..391T}
{Timokhin} A.~N.,  {Aharonian} F.~A.,    {Neronov} A.~Y.,  2004, \aap, 417, 391

\bibitem[\protect\citeauthoryear{{Vestrand}}{{Vestrand}}{1982}]{1982AJ.....87.%
1266V}
{Vestrand} W.~T.,  1982, \aj, 87, 1266

\bibitem[\protect\citeauthoryear{{Vogt} \& {En{\ss}lin}}{{Vogt} \&
  {En{\ss}lin}}{2005}]{2005A&A...434...67V}
{Vogt} C.,  {En{\ss}lin} T.~A.,  2005, \aap, 434, 67

\bibitem[\protect\citeauthoryear{{V{\"o}lk}, {Aharonian} \&
  {Breitschwerdt}}{{V{\"o}lk} et~al.}{1996}]{1996SSRv...75..279V}
{V{\"o}lk} H.~J.,  {Aharonian} F.~A.,    {Breitschwerdt} D.,  1996, Space
  Science Reviews, 75, 279

\bibitem[\protect\citeauthoryear{{Widrow}}{{Widrow}}{2002}]{2002RvMP...74..775%
W}
{Widrow} L.~M.,  2002, Reviews of Modern Physics, 74, 775

\bibitem[\protect\citeauthoryear{{Wolfe} \& {Melia}}{{Wolfe} \&
  {Melia}}{2007}]{2007arXiv0712.0187W}
{Wolfe} B.,  {Melia} F.,  2007, ArXiv:0712.0187

\end{thebibliography}
\bibliographystyle{mn2e}

\appendix

\section{Predicting the brightest and most-luminous nearby $\bgamma$-ray clusters}

\begin{table*}
\caption{\scshape The brightest IC/$\gamma$-ray clusters of the HIFLUGCS sample:$^{(1)}$}
\begin{tabular}{l c c  c c c c  c c c c }
\hline
\hline
& & & 
\multicolumn{4}{c}{IC emission, $E_\gamma> 10$~keV:} & 
\multicolumn{4}{c}{$\gamma$-ray emission, $E_\gamma> 100$~MeV:} \\
cluster name & $z$ & $M_{200}^{(2)}$ 
& $\F_\rmn{IC}^{(3)}$ (S1)        & $\F_\rmn{IC}^{(3)}$ (S2) 
& $\F_\rmn{IC}^{(3)}$ (S2, $B_3$) & $\F_\rmn{IC}^{(3)}$ (S3) 
& $\F_\gamma^{(4)}$ (S1)          & $\F_\gamma^{(4)}$ (S2)   
& $\F_\gamma^{(4)}$ (S2, $B_3$)   & $\F_\gamma^{(4)}$ (S3) \\
\hline
Ophiuchus  & 0.0280 &  2.32 &  3.43 &  1.77 &  2.40 &  2.26 &  9.11 &  5.75 &  5.95 &  8.49 \\
Fornax     & 0.0046 &  0.10 &  1.12 &  1.02 &  1.42 &  1.82 &  3.04 &  3.55 &  3.68 &  8.38 \\
Coma       & 0.0232 &  1.38 &  2.30 &  1.30 &  1.78 &  1.76 &  6.12 &  4.28 &  4.43 &  6.82 \\
A3627      & 0.0163 &  0.66 &  1.51 &  0.98 &  1.35 &  1.43 &  4.04 &  3.27 &  3.39 &  5.84 \\
Perseus    & 0.0183 &  0.77 &  1.52 &  0.96 &  1.32 &  1.38 &  4.08 &  3.20 &  3.31 &  5.57 \\
A3526      & 0.0103 &  0.27 &  0.98 &  0.75 &  1.04 &  1.21 &  2.65 &  2.56 &  2.65 &  5.21 \\
A1060      & 0.0114 &  0.30 &  0.96 &  0.71 &  0.99 &  1.13 &  2.57 &  2.43 &  2.51 &  4.86 \\
M49        & 0.0044 &  0.05 &  0.37 &  0.39 &  0.55 &  0.76 &  1.01 &  1.38 &  1.43 &  3.67 \\
AWM7       & 0.0172 &  0.43 &  0.72 &  0.50 &  0.69 &  0.77 &  1.94 &  1.70 &  1.76 &  3.22 \\
3C129      & 0.0223 &  0.66 &  0.81 &  0.53 &  0.72 &  0.77 &  2.18 &  1.76 &  1.82 &  3.13 \\
NGC4636    & 0.0037 &  0.03 &  0.20 &  0.24 &  0.34 &  0.50 &  0.56 &  0.87 &  0.90 &  2.52 \\
A1367      & 0.0216 &  0.41 &  0.41 &  0.29 &  0.40 &  0.45 &  1.10 &  0.98 &  1.02 &  1.88 \\
A0754      & 0.0528 &  1.87 &  0.67 &  0.36 &  0.49 &  0.47 &  1.78 &  1.18 &  1.22 &  1.79 \\
Triangulum & 0.0510 &  1.54 &  0.54 &  0.30 &  0.41 &  0.40 &  1.43 &  0.98 &  1.01 &  1.53 \\
NGC5846    & 0.0061 &  0.04 &  0.14 &  0.15 &  0.22 &  0.31 &  0.39 &  0.55 &  0.57 &  1.50 \\
\hline
\end{tabular}
\begin{quote} 
  {\scshape Notes:}\\ 
  (1) IC and $\gamma$-ray fluxes of clusters that are contained in the complete
  sample of the X-ray brightest clusters (HIFLUGCS, the HIghest X-ray FLUx
  Galaxy Cluster Sample, \citet{2002ApJ...567..716R}). We predict these fluxes
  using our cluster scaling relations for non-thermal observables defined in
  Table~\ref{tab:scaling}. The definition for our different models can be found
  in Table~\ref{tab:models}. The clusters are ordered according to their
  decreasing $\gamma$-ray flux in our model S3.\\
  (2) Mass contained within $R_{200}$ in units of $10^{15}
  h_{70}^{-1}~\rmn{M}_\odot$.\\
  (3) Predicted total (primary and secondary) IC flux in units of $10^{-5}
  \gamma \mbox{ cm}^{-2} \mbox{ s}^{-1} h_{70}^3$. If not otherwise mentioned,
  we use our magnetic parametrisation of $B_0=10\,\umu$G and
  $\alpha_B=0.5$. Our model with $B_3$ refers to a smaller central value for
  the magnetic field of $B_0=3\,\umu$G and yields a higher IC luminosity.\\
  (4) Predicted total (primary IC, secondary IC, pion decay) $\gamma$-ray
  flux in units of $10^{-9} \gamma \mbox{ cm}^{-2} \mbox{ s}^{-1} h_{70}^3$. If not 
  otherwise mentioned, we use our magnetic parametrisation of $B_0=10\,\umu$G
  and $\alpha_B=0.5$. Our model with $B_3$ refers to a smaller central value
  for the magnetic field of $B_0=3\,\umu$G which barely effects the
  $\gamma$-ray flux due to the dominant contribution from pion decay
  emission. \\
\end{quote}
\label{tab:flux}
\end{table*}

\begin{table*}
\caption{\scshape The most luminous IC/$\gamma$-ray clusters of the HIFLUGCS sample:$^{(1)}$}
\begin{tabular}{l c c  c c c c  c c c c }
\hline
\hline
& & & 
\multicolumn{4}{c}{IC emission, $E_\gamma> 10$~keV:} & 
\multicolumn{4}{c}{$\gamma$-ray emission, $E_\gamma> 100$~MeV:} \\
cluster name & $z$ & $M_{200}^{(2)}$ 
& $\L_\rmn{IC}^{(3)}$ (S1)        & $\L_\rmn{IC}^{(3)}$ (S2) 
& $\L_\rmn{IC}^{(3)}$ (S2, $B_3$) & $\L_\rmn{IC}^{(3)}$ (S3) 
& $\L_\gamma^{(4)}$ (S1)          & $\L_\gamma^{(4)}$ (S2)   
& $\L_\gamma^{(4)}$ (S2, $B_3$)   & $\L_\gamma^{(4)}$ (S3) \\
\hline
A2163      & 0.2010 &  3.71 & 12.64 &  6.00 &  8.10 &  7.28 &  3.35 &  1.93 &  1.99 &  2.65 \\
A3888      & 0.1510 &  2.55 &  7.14 &  3.63 &  4.92 &  4.58 &  1.89 &  1.17 &  1.21 &  1.71 \\
A1914      & 0.1712 &  2.43 &  6.62 &  3.39 &  4.60 &  4.31 &  1.76 &  1.10 &  1.14 &  1.61 \\
Ophiuchus  & 0.0280 &  2.32 &  6.16 &  3.18 &  4.32 &  4.06 &  1.64 &  1.03 &  1.07 &  1.52 \\
A3827      & 0.0980 &  1.96 &  4.77 &  2.54 &  3.46 &  3.31 &  1.27 &  0.83 &  0.86 &  1.25 \\
A0754      & 0.0528 &  1.87 &  4.45 &  2.39 &  3.25 &  3.12 &  1.18 &  0.78 &  0.81 &  1.19 \\
A1689      & 0.1840 &  1.76 &  4.06 &  2.21 &  3.00 &  2.90 &  1.08 &  0.72 &  0.75 &  1.11 \\
A3266      & 0.0594 &  1.70 &  3.83 &  2.10 &  2.85 &  2.77 &  1.02 &  0.69 &  0.71 &  1.06 \\
A2065      & 0.0721 &  1.67 &  3.74 &  2.05 &  2.79 &  2.71 &  0.99 &  0.67 &  0.69 &  1.04 \\
A2256      & 0.0601 &  1.56 &  3.36 &  1.87 &  2.54 &  2.49 &  0.90 &  0.61 &  0.63 &  0.96 \\
Triangulum & 0.0510 &  1.54 &  3.30 &  1.84 &  2.50 &  2.45 &  0.88 &  0.60 &  0.62 &  0.94 \\
A2142      & 0.0899 &  1.50 &  3.18 &  1.78 &  2.42 &  2.38 &  0.85 &  0.58 &  0.60 &  0.92 \\
A0644      & 0.0704 &  1.42 &  2.91 &  1.64 &  2.24 &  2.21 &  0.78 &  0.54 &  0.56 &  0.86 \\
Coma       & 0.0232 &  1.38 &  2.81 &  1.59 &  2.17 &  2.15 &  0.75 &  0.52 &  0.54 &  0.83 \\
A2029      & 0.0767 &  1.34 &  2.68 &  1.53 &  2.08 &  2.07 &  0.71 &  0.50 &  0.52 &  0.80 \\
\hline
\end{tabular}
\begin{quote} 
  {\scshape Notes:}\\ 
  (1) IC and $\gamma$-ray luminosities of clusters that are contained in the
  complete sample of the X-ray brightest clusters (HIFLUGCS, the HIghest X-ray
  FLUx Galaxy Cluster Sample, \citet{2002ApJ...567..716R}). We predict these
  luminosities using our cluster scaling relations defined in
  Table~\ref{tab:scaling}. The definition for our different models can be found
  in Table~\ref{tab:models}. The clusters are ordered according to their
  decreasing $\gamma$-ray luminosities in our model S3.\\
  (2) Mass contained within $R_{200}$ in units of $10^{15}
  h_{70}^{-1}~\rmn{M}_\odot$.\\
  (3) Predicted total (primary and secondary) IC luminosity in units of
  $10^{49} \gamma \mbox{ s}^{-1} h_{70}$. If not otherwise mentioned, we use
  our magnetic parametrisation of $B_0=10\,\umu$G and $\alpha_B=0.5$. Our model
  with $B_3$ refers to a smaller central value for the magnetic field of
  $B_0=3\,\umu$G and yields a higher IC luminosity.\\
  (4) Predicted total (primary IC, secondary IC, pion decay) $\gamma$-ray
  luminosity in units of $10^{46} \gamma \mbox{ s}^{-1} h_{70}$. If not
  otherwise mentioned, we use our magnetic parametrisation of $B_0=10\,\umu$G
  and $\alpha_B=0.5$. Our model with $B_3$ refers to a smaller central value
  for the magnetic field of $B_0=3\,\umu$G which barely effects the
  $\gamma$-ray luminosity due to the dominant contribution from pion decay
  emission. \\
\end{quote}
\label{tab:luminosity}
\end{table*}

High-energy $\gamma$-ray fluxes and luminosities a related by the simple
conversion formula,
\begin{eqnarray}
  \label{eq:flux_gamma}
  \mathcal{F}_\gamma &=& \frac{\mathcal{L}_\gamma}{4\pi\,D_\rmn{lum}^2} 
  \nonumber\\
  &=& 8.4\times 10^{-9} \frac{\gamma\,h_{70}^3}{\mbox{ cm}^2\mbox{ s}}\,
  \left(\frac{\mathcal{L_\gamma}}{10^{46} \mbox{ s}^{-1}\,h_{70}}\right)
 \left(\frac{D_\rmn{lum}}{100 \mbox{ Mpc}\,h_{70}^{-1}}\right)^{-2}.
\end{eqnarray}

\bsp

\label{lastpage}

\end{document}